\documentclass[letterpaper]{article}
\usepackage{aaai24}
\usepackage{times}
\usepackage{helvet}
\usepackage{courier}
\usepackage[hyphens]{url}
\usepackage{graphicx}
\def\UrlFont{\rm}
\usepackage{natbib}
\usepackage{caption}
\usepackage[edges]{forest}

\usepackage{algorithm}
\usepackage{algorithmic}
\usepackage{newfloat}
\usepackage{listings}

\graphicspath{ {./images/} }
\usepackage{comment}

\usepackage{amsmath}
\usepackage{booktabs}
\usepackage{tikz}
\usepackage{pgfplots}
\usepackage{pgfplotstable}
\usetikzlibrary{patterns}
\usepgfplotslibrary{colormaps}
\pgfplotsset{compat=1.18}
\usetikzlibrary{pgfplots.colormaps}
\usepackage{pgf-pie} 
\usepackage{xcolor,colortbl}
\usepackage{multirow}
\usepackage{xurl}
\usepackage{siunitx}

\usepackage{dcolumn}
\usepackage{makecell}

\usetikzlibrary{pgfplots.statistics}
\usetikzlibrary{matrix}
\usetikzlibrary{calc}

\newcolumntype{d}[1]{D{.}{.}{#1}}
\newcommand{\mc}[1]{\multicolumn{1}{c@{}}{#1}}
\newcommand{\gray}[1]{%
{\cellcolor{gray!\number\numexpr7*#1/12\relax}}%
}

\pgfplotstableread{
human_means	teen_model_cos_sims
3.642857142857143	0.49776632
3.4285714285714284	0.2935213
3.7857142857142856	0.32999068
3.857142857142857	0.39302292
3.9285714285714284	0.18831103
4.214285714285714	0.23293573
4.0	0.36074027
3.5	0.21640536
3.142857142857143	0.3973443
2.857142857142857	0.23759939
3.2857142857142856	0.3854057
3.857142857142857	0.3929243
3.0714285714285716	0.33241212
3.9285714285714284	0.40002567
3.642857142857143	0.336502
3.357142857142857	0.08974466
3.5	0.40053004
4.642857142857143	0.24468458
3.0	0.28670356
3.0	0.23142964
}\glovetable

\pgfplotstableread{
humanMeans	modelSims
3.642857142857143	0.4221193
3.4285714285714284	0.20580746
3.7857142857142856	0.30917248
3.857142857142857	0.1552194
3.9285714285714284	0.1384141
4.214285714285714	0.09316026
4.0	0.31598744
3.5	0.12220531
3.142857142857143	0.21071346
2.857142857142857	0.13994364
3.2857142857142856	0.3479868
3.857142857142857	0.25991142
3.0714285714285716	0.25239545
3.9285714285714284	0.1743945
3.642857142857143	0.24480602
3.357142857142857	0.06871239
3.5	0.28151867
4.642857142857143	0.22823809
3.0	0.22750334
3.0	0.18256256
}\fttable

\pgfplotstableread{
word	dim1	dim2	cluster
set	-0.515670120716095	7.4654412269592285	2
site	1.1870814561843872	6.6041059494018555	4
After	-1.6015654802322388	7.336176872253418	5
music	1.5307639837265015	5.994600772857666	6
series	0.3004136383533478	7.106053829193115	1
writing	-0.03145934268832207	6.649201393127441	1
issue	-0.6348258256912231	7.853283882141113	2
film	0.43850061297416687	7.592187881469727	1
issues	-0.9499551653862	8.349384307861328	2
cover	-0.6235719323158264	7.720577239990234	2
drive	1.9402052164077759	7.796849727630615	2
First	-1.2480802536010742	7.131420612335205	5
began	-1.838020920753479	8.1901216506958	2
summer	-1.0869334936141968	5.935375213623047	6
girl	1.736774206161499	9.953536033630371	0
period	-1.85856294631958	5.648742198944092	2
London	5.580996990203857	7.111791133880615	5
became	-1.861334204673767	8.260557174682617	2
network	1.5373821258544922	6.7393083572387695	4
track	1.4486140012741089	5.883724689483643	6
April	-1.2404193878173828	5.815863132476807	5
weekend	-0.9777891635894775	6.016381740570068	6
hot	-0.3806301951408386	9.861957550048828	0
band	1.9085607528686523	5.3584208488464355	6
policy	-0.04317816346883774	8.329731941223145	2
June	-1.243483543395996	5.801431179046631	5
2005	-1.7196489572525024	5.539450645446777	5
internet	1.1541500091552734	6.576427459716797	4
July	-1.2223405838012695	5.825107097625732	5
Internet	1.2917990684509277	6.6455841064453125	4
sites	1.3212860822677612	6.757297039031982	4
college	2.8979151248931885	9.010529518127441	6
scene	0.1970675140619278	7.657492637634277	1
led	-1.8052383661270142	8.551970481872559	2
setting	-0.5428237915039062	7.499840259552002	2
Facebook	0.8389956951141357	6.419410705566406	4
August	-1.247389554977417	5.797590732574463	5
sex	1.2838644981384277	10.421894073486328	0
target	-0.09166861325502396	8.969947814941406	2
driving	1.9193661212921143	7.781595706939697	2
spring	-1.1613582372665405	5.877486705780029	6
club	2.939605712890625	8.714784622192383	6
Review	-0.6369113922119141	7.0234293937683105	5
competition	0.7396835088729858	8.596961975097656	2
pressure	-1.1626232862472534	9.244462013244629	2
writer	-0.0718204453587532	6.718642234802246	1
Share	-0.5550109148025513	6.919901371002197	5
influence	-0.5576834678649902	8.55848503112793	2
discovered	-1.6090556383132935	8.053773880004883	2
battle	0.5572877526283264	8.84163761138916	2
volume	-0.35768112540245056	7.639710903167725	2
session	0.9101491570472717	5.809048175811768	6
traffic	1.7005658149719238	7.316612720489502	4
Twitter	0.7573528289794922	6.370634078979492	4
rock	1.346978783607483	8.637770652770996	6
rise	-0.7944886684417725	8.634143829345703	2
North	4.887072563171387	7.239392280578613	5
posting	0.6582100987434387	6.282451629638672	4
profile	0.5471826195716858	6.892401695251465	4
caught	-0.10091274976730347	9.632558822631836	2
trouble	-1.5604671239852905	9.256746292114258	2
opinion	-0.47495540976524353	8.011069297790527	2
College	2.8489937782287598	8.702545166015625	5
interview	-0.2881815731525421	6.4939141273498535	4
covers	-0.8454203009605408	7.643940448760986	2
shooting	0.3487464487552643	8.29271411895752	2
drugs	1.01663339138031	11.839643478393555	6
Paris	5.576963424682617	6.803076267242432	5
radio	1.6759538650512695	6.3014912605285645	4
channel	1.7838865518569946	6.606042385101318	4
initial	-1.5580849647521973	7.395365238189697	2
violence	-0.4353300929069519	10.779793739318848	1
crime	-0.3201502561569214	10.997608184814453	1
drug	0.9822686314582825	11.843289375305176	6
identify	-1.133348822593689	7.762502670288086	2
ended	-1.8217028379440308	8.401737213134766	2
magazine	-0.3215331435203552	7.757338523864746	6
Set	-0.571745753288269	7.317437171936035	5
feedback	0.2411956936120987	6.307758808135986	4
drivers	2.1176974773406982	7.668114185333252	6
caused	-1.6943554878234863	8.812439918518066	2
charges	-0.6867741346359253	11.177196502685547	2
dangerous	-1.384606957435608	10.159777641296387	2
debt	-0.6094744801521301	10.007994651794434	2
youth	2.51713490486145	9.503143310546875	6
Team	2.904348134994507	8.484618186950684	5
launched	-1.6369328498840332	8.13855266571045	2
attempt	-1.8598779439926147	8.920763969421387	2
forces	-1.0242451429367065	9.354522705078125	2
chapter	0.17109714448451996	7.041351318359375	1
scenes	0.25374922156333923	7.448976993560791	1
struggle	-1.664272427558899	9.151076316833496	2
plot	0.1691100001335144	7.29750919342041	1
hopes	-1.6561232805252075	9.272082328796387	2
High	1.5482820272445679	8.511550903320312	5
causing	-1.5079841613769531	9.015420913696289	2
pop	0.37822890281677246	9.005631446838379	6
Second	-1.2392905950546265	7.124819755554199	5
contest	0.8961983323097229	8.613511085510254	6
Summer	-1.0401887893676758	5.97213888168335	5
wild	-1.3750112056732178	10.214316368103027	2
listening	1.4468982219696045	6.050866603851318	4
string	2.47865891456604	6.422814846038818	6
mass	-0.702703058719635	8.815171241760254	2
alcohol	0.8296731114387512	11.899054527282715	6
murder	-0.2928328812122345	11.045963287353516	1
Sam	7.182861328125	7.843403339385986	3
deals	-0.4111148416996002	8.348048210144043	2
Martin	7.107439041137695	6.284374237060547	3
1999	-1.7921808958053589	5.560842990875244	5
tone	-0.4244915246963501	7.738099575042725	2
Seattle	5.2247138023376465	7.39661169052124	5
fuck	1.4934784173965454	10.445813179016113	0
gay	1.0255274772644043	10.517362594604492	0
acts	-0.4513469636440277	10.80286979675293	2
charged	-0.6696986556053162	11.158543586730957	2
Ryan	7.531130313873291	7.391320705413818	3
Author	-0.466625452041626	6.81675386428833	5
drinking	0.8013964891433716	11.857748031616211	6
Toronto	5.444009304046631	7.535472869873047	5
fucking	1.5584810972213745	10.543868064880371	0
drama	0.3481285870075226	7.768414497375488	1
Young	2.514622449874878	9.237297058105469	5
strike	-0.2463339865207672	9.154847145080566	2
exploring	-1.4987086057662964	7.906158924102783	2
brief	-1.5702499151229858	7.318613529205322	2
gym	2.966524124145508	8.43993854522705	6
concert	1.7788902521133423	5.4057841300964355	6
hitting	-0.23396694660186768	9.070635795593262	2
Rock	1.6956745386123657	8.440710067749023	5
mood	-0.33932945132255554	7.9354071617126465	2
guitar	2.209832191467285	5.637692451477051	6
studying	-1.4047985076904297	7.807203769683838	2
wave	-0.7047123908996582	8.669248580932617	2
Queen	5.324003219604492	8.096449851989746	5
discovery	-1.4992825984954834	8.065242767333984	2
sessions	1.0149250030517578	5.832503318786621	6
ass	1.616171956062317	10.532413482666016	0
storm	-0.4310663938522339	9.030349731445312	2
extreme	-1.5092984437942505	10.030706405639648	2
bands	1.878172755241394	5.398934364318848	6
flash	0.7347272038459778	9.281904220581055	2
initially	-1.883320689201355	7.7182936668396	2
drives	1.8535354137420654	7.798276424407959	2
quotes	0.3560344874858856	6.636196613311768	2
1997	-1.8521398305892944	5.557361125946045	5
interviews	-0.10333389788866043	6.659450531005859	4
lyrics	1.674237847328186	5.783667087554932	6
garage	2.5001614093780518	7.597555637359619	6
Model	1.851590633392334	8.761198043823242	5
anger	-0.9472339153289795	9.973061561584473	2
Nick	7.381702899932861	7.386249542236328	3
frequency	-0.6255266070365906	8.249635696411133	2
crash	-0.7307380437850952	9.258784294128418	2
Jason	7.474701881408691	7.373117923736572	3
Sydney	5.555944919586182	7.415538787841797	5
York	5.343942165374756	7.434137344360352	5
grew	-1.7836928367614746	8.393890380859375	2
Taylor	6.97389554977417	6.037056922912598	3
YouTube	1.2316564321517944	6.450803279876709	4
driven	1.7425544261932373	7.893178462982178	2
trends	-0.3692163825035095	8.206473350524902	2
Jordan	6.799191474914551	7.231295108795166	3
editing	0.13555601239204407	6.538080215454102	4
gained	-1.6551754474639893	8.42463493347168	2
genre	0.8331546783447266	7.325318813323975	1
Girl	2.1232423782348633	9.014333724975586	5
Victoria	5.906983852386475	8.010100364685059	3
Writing	-0.14261461794376373	6.685513973236084	1
clubs	2.745417833328247	8.804953575134277	6
drove	1.6044377088546753	8.017607688903809	2
dialogue	0.07260320335626602	7.332090854644775	1
Drive	2.1348724365234375	7.784229278564453	5
circuit	2.3503077030181885	6.244206428527832	6
contacts	0.5156682729721069	6.914973258972168	4
Future	1.9296318292617798	8.690010070800781	5
attempts	-1.9038426876068115	8.914375305175781	2
Amy	7.202460289001465	8.48381233215332	3
romance	0.460642009973526	7.665650367736816	1
encounter	-1.3845025300979614	8.49547290802002	2
pricing	-0.05988452956080437	8.303070068359375	2
essay	0.02715481072664261	6.646301746368408	1
struggling	-1.731634259223938	9.1250581741333	2
sexual	0.9351603388786316	10.55986213684082	0
Wilson	6.992569446563721	5.770025730133057	3
intense	-1.4177881479263306	10.050874710083008	2
crack	0.09690134972333908	9.620837211608887	2
singer	1.8324593305587769	5.54767370223999	6
arrested	-0.5012209415435791	11.298396110534668	2
smoking	0.8777057528495789	12.023130416870117	6
virus	-0.6592264175415039	10.252520561218262	2
Dean	7.156340599060059	7.744968891143799	3
Girls	2.1927783489227295	9.115472793579102	5
Hot	1.3773773908615112	8.596341133117676	5
Ontario	5.437703609466553	7.529924392700195	5
makeup	1.479743242263794	9.479581832885742	6
teens	2.513896942138672	9.57681655883789	6
Rachel	7.210908889770508	8.44107437133789	3
Rob	7.326513290405273	7.437277793884277	3
bass	2.188936471939087	5.723113536834717	6
solo	1.8220843076705933	5.379453659057617	6
struck	-0.32237526774406433	9.082576751708984	2
Columbia	5.248015880584717	7.208462238311768	5
teen	2.5312042236328125	9.61530590057373	6
Beauty	1.7393425703048706	9.134322166442871	5
confusion	-1.5126962661743164	9.4330472946167	2
overnight	-0.36338597536087036	9.619879722595215	2
Policy	1.068013310432434	8.133787155151367	5
Instagram	0.8682056665420532	6.32123327255249	4
debut	-0.724669337272644	7.230660438537598	6
League	3.1621124744415283	8.759713172912598	5
Lewis	6.925396919250488	5.863185882568359	3
Edward	6.876822471618652	7.9101362228393555	3
Anonymous	0.17106013000011444	10.413924217224121	5
threats	-0.6033539175987244	10.479498863220215	2
identifying	-1.2158112525939941	7.699191570281982	2
arrest	-0.5626342296600342	11.225083351135254	2
killer	-0.19119250774383545	10.487281799316406	2
texts	0.5040776133537292	6.634045124053955	4
popularity	-0.9557173848152161	8.655482292175293	2
tension	-1.2780059576034546	9.276922225952148	2
jeans	2.005030393600464	10.662440299987793	0
downloaded	1.089460015296936	6.048612594604492	4
violent	-1.2509270906448364	10.353654861450195	2
porn	1.2980365753173828	10.396537780761719	0
D.C.	5.203773021697998	7.388606548309326	5
Latin	5.155158519744873	6.859898090362549	5
naked	1.761918544769287	10.313752174377441	0
desperate	-1.6069458723068237	9.664817810058594	2
novels	0.40648353099823	7.115704536437988	1
Student	2.5333313941955566	8.636013984680176	5
Pink	2.2099609375	9.059704780578613	5
attempted	-1.918197751045227	8.73930549621582	2
charts	-0.40245336294174194	8.21668815612793	2
Poland	5.516726493835449	6.654909610748291	5
cock	1.533636450767517	10.540284156799316	0
Grant	7.203738689422607	6.016425132751465	3
frustration	-1.3683569431304932	9.569892883300781	2
struggles	-1.6145159006118774	9.08593463897705	2
Tech	2.775792360305786	8.015633583068848	5
Jessica	6.960966110229492	8.476055145263672	3
profiles	0.5836300849914551	6.899625301361084	4
controversial	-1.690429925918579	10.002985954284668	2
girlfriend	1.640589952468872	10.106111526489258	0
Justin	7.4656596183776855	7.308030128479004	3
Angel	5.267922401428223	8.481313705444336	3
Clark	7.007086753845215	5.789462089538574	3
Oxford	5.594832897186279	7.194594383239746	5
Morgan	7.05769681930542	6.257772445678711	3
Aaron	7.419437885284424	7.490751266479492	3
Hamilton	6.733970642089844	6.076906681060791	3
GPS	2.414379358291626	7.228691101074219	4
boyfriend	1.7807799577713013	10.009841918945312	0
isolated	-1.2824488878250122	9.86860179901123	2
Third	-1.2469875812530518	7.13403844833374	5
confusing	-1.6847001314163208	9.706513404846191	2
squad	3.0725924968719482	8.769441604614258	6
acid	0.740236222743988	11.721631050109863	6
prospects	0.07363266497850418	7.7887492179870605	2
Chelsea	6.633694171905518	8.153081893920898	3
Caribbean	5.131940841674805	7.025752544403076	5
Nashville	5.1926589012146	7.492007732391357	5
chemistry	0.4750826358795166	7.7766032218933105	6
junior	3.020932912826538	9.09406566619873	6
terrorism	-0.28027546405792236	10.893645286560059	1
Dublin	5.614186763763428	6.912782192230225	5
Avenue	3.5763401985168457	7.302999496459961	5
Movies	0.8366883993148804	7.721997261047363	1
chapters	0.20570074021816254	6.994504451751709	1
Lane	3.849295139312744	7.125677585601807	3
Bell	6.998082160949707	5.776137351989746	3
gang	-0.1885758489370346	11.035408973693848	6
Guy	7.376518726348877	7.345430374145508	3
Mississippi	5.347641944885254	7.2621283531188965	5
scholarship	2.886554002761841	9.01632022857666	6
Ross	6.860069274902344	5.866782188415527	3
briefly	-1.8718233108520508	7.737058639526367	2
shorts	2.0112459659576416	10.690085411071777	6
Battle	0.6921042203903198	8.7087984085083	5
Madison	6.7217278480529785	6.429781913757324	3
dynamics	-0.43683770298957825	8.49538803100586	2
Youth	2.455296039581299	9.202433586120605	5
encountered	-1.534620761871338	8.411788940429688	2
Montreal	5.379318714141846	7.407026290893555	5
suicide	-0.3086305260658264	10.92790412902832	2
Israeli	5.450664520263672	6.6519012451171875	5
taxi	2.257866859436035	7.660242080688477	6
nail	0.6101033687591553	9.47640609741211	2
Sean	7.498621940612793	7.36361837387085	3
mod	1.9936878681182861	6.19374418258667	6
facebook	0.8346275091171265	6.439209938049316	4
Dawn	7.139562606811523	8.536436080932617	3
tactics	-0.0770685151219368	8.515069007873535	2
passionate	-1.4006584882736206	10.112582206726074	2
heated	-0.7056434750556946	9.86781120300293	2
launching	-1.2373013496398926	8.094593048095703	2
mall	3.228307008743286	7.498879909515381	6
overwhelming	-1.631150245666504	9.75705623626709	2
casting	0.4072571098804474	8.014052391052246	2
1970s	-1.9105818271636963	5.478772163391113	6
strikes	-0.4110791087150574	9.099055290222168	2
Louis	6.750677585601807	7.864815711975098	3
Band	1.9770472049713135	5.354552268981934	5
nails	1.111788272857666	9.467833518981934	6
subsequently	-1.9132461547851562	7.747147083282471	2
battles	0.4967805743217468	8.831337928771973	2
addiction	0.8435521125793457	11.664320945739746	6
surf	1.193277359008789	6.7299065589904785	4
Carter	7.245334148406982	5.981054782867432	3
Boys	2.3060059547424316	9.062408447265625	5
Montana	5.099348545074463	7.241215229034424	5
Ashley	6.76935338973999	8.1748046875	3
1980s	-1.9563169479370117	5.4318671226501465	6
Harris	6.97452449798584	5.79645299911499	3
panic	-1.3034435510635376	9.638666152954102	2
teenagers	2.4873335361480713	9.656342506408691	6
wires	2.525106430053711	6.393955230712891	6
1960s	-1.9177780151367188	5.467860698699951	6
MB	4.738865852355957	7.54296875	5
Northern	4.992700099945068	7.224699020385742	5
pins	2.4526777267456055	6.330661296844482	6
momentum	-0.8871483206748962	8.691553115844727	2
Tyler	7.520514488220215	7.1444549560546875	3
essays	0.07483823597431183	6.682941436767578	1
conviction	-0.7988107800483704	11.145172119140625	2
modeling	1.382246494293213	9.245428085327148	6
underground	2.024517297744751	7.096439838409424	6
assault	-0.6827914118766785	10.882946968078613	2
wasted	-1.6433441638946533	9.657288551330566	2
clips	1.3101873397827148	6.415084362030029	4
Sara	7.139336109161377	8.56273078918457	3
Atlantic	5.006002902984619	7.183614730834961	5
anime	1.028066635131836	7.465398788452148	1
bitch	1.5140804052352905	10.461599349975586	0
downloading	1.0591801404953003	6.134738445281982	4
Rate	-0.5799371004104614	6.9264912605285645	5
twitter	0.743279755115509	6.437331676483154	4
Junior	2.9149837493896484	9.08387279510498	5
Danny	7.493976593017578	7.495723724365234	3
explored	-1.525422215461731	7.982258319854736	2
vocals	1.948453664779663	5.579819202423096	6
pussy	1.595376968383789	10.461008071899414	0
Kennedy	7.2633891105651855	5.891439914703369	3
Marie	6.898316383361816	8.599225997924805	3
nationwide	1.3249074220657349	6.96681022644043	2
Cameron	7.354238986968994	6.69079065322876	3
lit	0.7869617342948914	9.38625717163086	6
Angels	4.997651100158691	8.253954887390137	5
Lawrence	7.0736775398254395	6.0483503341674805	3
Nicole	6.9349589347839355	8.613565444946289	3
Bella	6.852281093597412	8.292092323303223	3
Blues	4.6157612800598145	7.685429096221924	5
Adams	7.034509658813477	5.856955051422119	3
Thompson	7.017807960510254	5.847781658172607	3
tunnel	2.0711967945098877	7.045695781707764	2
Lauren	6.892832279205322	8.354775428771973	3
duo	1.765441656112671	5.462747573852539	6
marijuana	0.9933597445487976	12.00352954864502	6
shy	1.3577194213867188	10.073477745056152	2
trio	1.761597752571106	5.453478813171387	6
rage	-0.9259660243988037	9.935646057128906	2
AA	3.203601121902466	8.963323593139648	6
surveys	-0.11520237475633621	6.908487796783447	2
Dell	3.5140445232391357	7.52032995223999	3
Lebanon	5.616683006286621	6.597269535064697	5
Perry	7.1734466552734375	5.968142032623291	3
Polish	5.382814884185791	6.622357368469238	5
Speed	1.9708665609359741	7.796619415283203	5
startup	-0.894683837890625	7.4596333503723145	6
deadly	-1.1021031141281128	10.142857551574707	2
explores	-1.340407371520996	7.858429431915283	2
gaining	-1.1943024396896362	8.357348442077637	2
excessive	-1.4394258260726929	9.828425407409668	2
downloads	1.1010591983795166	6.082357406616211	4
brutal	-1.386580228805542	10.132431030273438	2
outlet	2.4312918186187744	6.9786224365234375	6
dash	0.5337536334991455	8.993370056152344	6
Claire	6.890606880187988	8.411433219909668	3
bullet	0.2408396154642105	8.582632064819336	2
discovering	-1.523546576499939	7.91663122177124	2
struggled	-1.810226321220398	8.897504806518555	2
Stewart	6.902397155761719	5.792469024658203	3
Broadway	3.752204418182373	7.277836799621582	5
butt	1.7350225448608398	10.58080768585205	0
Summary	-0.7645938396453857	7.083421230316162	5
protests	-0.7306178212165833	10.645302772521973	2
Pop	0.5358049869537354	8.815805435180664	5
Diamond	5.783592700958252	8.330997467041016	5
Front	1.0507359504699707	8.563716888427734	5
switches	2.370163679122925	6.425447940826416	2
Channel	1.954030990600586	6.769407749176025	5
responding	0.4348406493663788	6.310769081115723	2
polish	1.1501109600067139	9.468743324279785	6
timeline	0.10921377688646317	6.8124918937683105	4
invasion	-0.5952149033546448	10.623613357543945	2
Bristol	5.78921365737915	7.208877086639404	5
streaming	1.1991395950317383	6.202494144439697	4
conspiracy	-0.31081876158714294	10.851766586303711	1
MP3	1.3821383714675903	6.05211067199707	4
barriers	-0.7322186827659607	9.719889640808105	2
motorcycle	2.485300302505493	7.681291580200195	6
Beth	7.225675582885742	8.530077934265137	3
Milan	5.639184474945068	6.919264793395996	5
Brandon	7.553769111633301	7.344986438751221	3
Los	5.454755783081055	7.578206539154053	5
attitudes	-0.2630642056465149	8.246565818786621	2
Arab	5.465198993682861	6.597917556762695	5
Secret	1.9083914756774902	8.683034896850586	5
gig	1.8348594903945923	5.311829090118408	6
pills	1.10658597946167	11.69863510131836	0
outlook	-0.28289297223091125	8.092432022094727	2
amp	2.2280383110046387	5.949895858764648	6
Dick	7.361413955688477	7.3824920654296875	3
Metal	2.612582206726074	7.669220924377441	5
soundtrack	1.4907581806182861	5.910619735717773	1
Popular	1.1278246641159058	8.376619338989258	5
Gay	1.0875155925750732	10.470054626464844	5
Crystal	6.000211238861084	8.453766822814941	3
Metro	3.3231465816497803	7.284066677093506	5
explosion	-0.7987359166145325	9.251973152160645	2
mindset	-0.2649138569831848	8.30898666381836	2
Volume	-0.2942894697189331	7.4377875328063965	5
armed	-0.5771980285644531	10.66641616821289	2
1990s	-1.9010319709777832	5.4905171394348145	6
mirrors	2.1301259994506836	8.680359840393066	2
tattoo	1.7332830429077148	9.692853927612305	0
gauge	-0.9052923321723938	7.912356853485107	2
Neil	7.42106819152832	7.173694610595703	3
teenager	2.4946322441101074	9.67001724243164	6
Focus	1.0993925333023071	8.40559196472168	5
emerged	-1.6933867931365967	8.357718467712402	2
Episode	0.6909993290901184	8.022080421447754	5
bust	0.2567322254180908	9.794425964355469	2
tweets	0.4719414710998535	6.483525276184082	4
Mitchell	7.006679058074951	5.9558868408203125	3
Diana	6.733867168426514	8.80898380279541	3
speeds	1.6313151121139526	7.60463809967041	2
troubles	-1.5376091003417969	9.18810749053955	2
spotlight	0.9850410223007202	8.934269905090332	2
touring	1.8006393909454346	5.327576637268066	6
hardcore	1.3511919975280762	9.576881408691406	6
shoots	0.3734203279018402	8.38082504272461	2
Genesis	3.443995237350464	7.97696590423584	5
crush	0.06623994559049606	9.58432674407959	2
Minneapolis	5.1930670738220215	7.450599670410156	5
Login	-0.5233785510063171	6.8931145668029785	5
visibility	1.2576369047164917	7.232250213623047	2
pole	2.5478289127349854	6.413242816925049	6
Starbucks	3.1909046173095703	7.485067844390869	5
Jesse	7.44389009475708	7.554426670074463	3
amid	-1.6852078437805176	9.180707931518555	2
filming	0.39619919657707214	7.754863739013672	4
peers	2.7024779319763184	9.55564022064209	6
Skype	0.8576240539550781	6.509487628936768	4
Ash	6.430340766906738	8.458663940429688	3
surfing	1.1343164443969727	6.734570026397705	4
Brazilian	5.274980068206787	6.847179412841797	5
demons	0.5735588073730469	10.165555000305176	1
branding	1.0337245464324951	7.301589488983154	4
buzz	-1.0341110229492188	8.791088104248047	2
analyzing	-1.2933008670806885	7.710622310638428	2
Megan	7.093794345855713	8.217462539672852	3
sensation	-1.2838090658187866	9.031932830810547	2
identifies	-1.2899117469787598	7.81865930557251	2
Troy	7.420050621032715	7.248488426208496	3
pill	1.1304478645324707	11.561301231384277	0
Spencer	7.327332973480225	6.8367390632629395	3
Storm	2.2804691791534424	8.095702171325684	5
Holly	6.5559468269348145	8.488606452941895	3
Driver	2.1505653858184814	7.71815299987793	5
cab	2.1788527965545654	7.621095180511475	6
Casey	7.42961311340332	7.660216808319092	3
boom	-0.7690737247467041	8.830459594726562	2
openly	0.9543051719665527	10.552508354187012	2
Amber	6.284910202026367	8.549675941467285	3
Evans	7.127242565155029	5.836758613586426	3
flooding	-0.6750672459602356	9.484335899353027	2
velocity	-0.582310676574707	8.659740447998047	2
Electric	2.740250825881958	7.455737590789795	5
Eagles	3.332758903503418	8.43239974975586	5
se	1.734864354133606	7.554797172546387	6
vampire	0.6545213460922241	10.18150806427002	1
quarterback	3.178248167037964	8.764806747436523	6
flames	-0.4947913885116577	9.753683090209961	2
sexuality	0.9861112236976624	10.603960990905762	0
Colombia	5.258849620819092	6.901399612426758	5
Competition	0.8822016716003418	8.62359619140625	5
80s	-1.9511983394622803	5.4317307472229	6
bang	-0.06918136030435562	9.35928726196289	2
L.A.	5.303832054138184	7.58465051651001	5
blonde	1.6902860403060913	10.001372337341309	0
Pearl	5.903902053833008	8.505462646484375	5
Midwest	4.992227554321289	7.255669116973877	5
lesbian	1.1232753992080688	10.470837593078613	0
amateur	2.051861047744751	9.54914665222168	6
SMS	0.7952590584754944	6.58574914932251	4
crust	0.532029390335083	9.232355117797852	6
bullying	-0.4495309591293335	10.759373664855957	2
drilling	0.5568474531173706	9.428409576416016	2
Collins	6.932613372802734	5.806398391723633	3
cult	0.4791143238544464	10.41358757019043	6
Hungary	5.567738056182861	6.608589172363281	5
Zone	2.519108295440674	7.978009223937988	5
Campus	2.832029104232788	8.256097793579102	5
suspects	-0.4666968882083893	11.204422950744629	2
warnings	-0.9771276712417603	9.913928985595703	2
storyline	0.2388833463191986	7.2925028800964355	1
OMG	1.1054580211639404	7.669365406036377	5
headphones	1.855315923690796	6.070794105529785	6
Turner	7.150083065032959	5.771897792816162	3
contests	0.859477162361145	8.657909393310547	6
Bailey	7.018565654754639	6.013561725616455	3
surge	-0.9550570845603943	8.70347785949707	2
Shannon	7.1600661277771	8.15442943572998	3
freshman	3.0699942111968994	9.043977737426758	6
Crime	-0.06456225365400314	10.942342758178711	1
dominated	-1.170354962348938	8.64172649383545	2
dealing	-0.972183883190155	8.370627403259277	2
Rio	5.237560749053955	7.083456039428711	5
nude	1.656351923942566	10.315637588500977	0
gateway	2.30983304977417	7.030130863189697	2
tossed	-0.16485683619976044	9.417512893676758	2
Written	-0.6377614736557007	6.976243495941162	5
Marcus	7.299820423126221	7.355172634124756	3
Tracy	6.929311752319336	8.723706245422363	3
dangers	-0.9030493497848511	9.99880599975586	2
earrings	1.8264645338058472	9.614407539367676	6
zombies	0.4045861065387726	10.268287658691406	1
wiring	2.5457887649536133	6.4000067710876465	6
manga	0.749694287776947	7.4111785888671875	1
finals	2.575827121734619	8.819846153259277	6
glitter	1.3941988945007324	9.410542488098145	6
breeze	-0.4031873941421509	8.851303100585938	2
alerts	1.198922038078308	7.31207799911499	2
Teen	2.418552875518799	9.32279109954834	5
Dylan	7.558283805847168	7.418463706970215	3
Prague	5.563464164733887	6.658303260803223	5
Initially	-1.7014210224151611	7.402343273162842	5
workouts	3.0120439529418945	8.434676170349121	6
Jamaica	5.248155117034912	7.149758338928223	5
FM	2.0033812522888184	6.546253204345703	5
lengths	2.397857904434204	6.4515700340271	2
troubled	-1.603181004524231	9.357460021972656	2
Olympic	2.748899459838867	8.571907043457031	5
pressures	-1.1744880676269531	9.221966743469238	2
Twilight	3.8090202808380127	8.189900398254395	1
rivals	0.6770391464233398	8.614373207092285	2
Natalie	6.770840644836426	8.56358528137207	3
fucked	1.488039493560791	10.41602611541748	0
temptation	-0.07578107714653015	9.996081352233887	2
Models	1.66508150100708	8.821112632751465	5
crashes	-0.7497449517250061	9.239603042602539	2
1930s	-1.9320480823516846	5.464059352874756	6
wrestling	1.1775404214859009	9.106012344360352	6
comparisons	-0.14177775382995605	8.051236152648926	2
hooked	0.11744008958339691	9.66313362121582	2
wholesale	-0.0494551919400692	8.408666610717773	2
encounters	-1.374513864517212	8.413731575012207	2
Drama	0.9017835259437561	7.7933549880981445	1
rookie	3.0365309715270996	8.947075843811035	6
zombie	0.46686840057373047	10.214203834533691	1
Attack	0.9552604556083679	8.498147010803223	5
wired	2.5853636264801025	6.5037617683410645	6
protagonist	0.25001686811447144	7.264151096343994	1
bullets	0.2042466253042221	8.567399978637695	2
U.K.	5.643860340118408	7.136078834533691	5
demon	0.6809009313583374	10.242924690246582	1
Tina	6.897552490234375	8.753600120544434	3
tattoos	1.7039867639541626	9.608349800109863	0
Aurora	5.906402587890625	8.154768943786621	5
Phillips	6.915017127990723	5.675823211669922	3
bout	0.7175881266593933	8.963193893432617	2
rant	-0.7199734449386597	10.125639915466309	2
rebels	-0.5808756351470947	10.547686576843262	6
curb	-0.011979712173342705	9.575541496276855	2
Blair	7.310119152069092	6.521139621734619	3
fierce	-1.287864327430725	10.174983024597168	2
70s	-1.9429888725280762	5.447494983673096	6
Driving	2.002760410308838	7.762572765350342	5
hooks	2.509096622467041	6.269165515899658	6
Rings	2.929041862487793	7.798525333404541	5
mp3	1.347398281097412	6.0125412940979	4
breasts	1.8058360815048218	10.664892196655273	0
peaks	-0.7604050636291504	8.246247291564941	2
stereo	1.9057255983352661	6.076085567474365	6
doubts	-1.6252774000167847	9.245856285095215	2
Queens	5.2809367179870605	7.643073558807373	5
cigarettes	0.9753963947296143	12.025162696838379	6
youtube	1.2413461208343506	6.340014457702637	4
fled	-0.47341927886009216	9.29998779296875	2
bra	1.9436976909637451	10.664836883544922	0
fifteen	3.0720691680908203	10.084141731262207	6
Racing	2.5348377227783203	7.953935623168945	5
Mall	3.3019113540649414	7.561734199523926	5
Tara	6.753389835357666	8.533178329467773	3
headaches	-1.5779225826263428	9.266324996948242	2
salon	1.5294464826583862	9.62729263305664	6
guitars	2.2626471519470215	5.702483177185059	6
Travis	7.594589710235596	7.3212714195251465	3
trilogy	0.294286847114563	7.14105749130249	1
Brighton	5.805403709411621	7.21407413482666	5
Setting	-0.5356533527374268	7.342451572418213	5
underwear	1.9724451303482056	10.671870231628418	0
make-up	1.4612739086151123	9.452996253967285	6
collision	-0.7159173488616943	9.219710350036621	2
hormones	1.191973090171814	11.246381759643555	0
singers	1.8311269283294678	5.4795823097229	6
acne	1.1683636903762817	11.046956062316895	0
lust	0.25583532452583313	10.278554916381836	0
Films	0.6848041415214539	7.768421649932861	1
murders	-0.3119506239891052	11.066014289855957	1
battling	0.43665748834609985	8.825370788574219	2
fatal	-1.085618019104004	9.971750259399414	2
Reagan	7.364276885986328	6.034515857696533	3
Traffic	1.860669493675232	7.414919853210449	5
Savannah	5.614990711212158	7.496518135070801	3
Welsh	5.8858747482299805	7.063529014587402	5
supernatural	0.6452709436416626	10.235404968261719	1
peer	2.60440731048584	9.584836959838867	2
Started	-1.5623953342437744	7.3523688316345215	5
Allison	7.033351898193359	8.258152961730957	3
weed	0.9754534959793091	12.082306861877441	6
popped	0.11849667876958847	9.185894012451172	2
punk	1.4881685972213745	9.212294578552246	6
kissing	1.580512285232544	9.950446128845215	0
Tiffany	6.837708473205566	8.405011177062988	3
ripped	-0.1718892753124237	9.442399978637695	2
narrator	0.18862318992614746	7.067880153656006	1
circuits	2.3920862674713135	6.342865943908691	6
Mercedes	6.360019207000732	7.990510940551758	3
homosexuality	0.902644693851471	10.630146026611328	0
Teens	2.4557652473449707	9.43835163116455	5
mob	-0.14911510050296783	10.998481750488281	2
Lindsay	7.127162456512451	8.069799423217773	3
Adrian	7.247382164001465	7.429537296295166	3
facial	1.535139560699463	9.704387664794922	0
automotive	2.577531337738037	7.536273956298828	6
Sullivan	7.113706588745117	5.983947277069092	3
rap	1.7355889081954956	6.274681568145752	6
ups	-0.7193036079406738	8.328154563903809	2
Bass	2.4032294750213623	5.565116882324219	5
Quinn	7.333251953125	6.960326194763184	3
Mirror	2.2713265419006348	8.536482810974121	5
Uber	2.2718636989593506	7.582265853881836	5
Samantha	6.713318347930908	8.451900482177734	3
aggression	-0.6711550354957581	10.626084327697754	2
prose	0.07636085152626038	6.755042552947998	1
Wireless	2.5648491382598877	7.211182594299316	5
Gibson	6.9742913246154785	5.861374855041504	3
portrayed	-0.03795647248625755	7.505394458770752	2
Hughes	7.191684246063232	5.715543746948242	3
90s	-1.884838581085205	5.493642807006836	6
subway	3.015655755996704	7.275599479675293	6
Czech	5.418050765991211	6.572897911071777	5
Connection	2.47880482673645	7.229157447814941	5
Chloe	6.6543965339660645	8.369179725646973	3
committing	-0.338011771440506	10.947943687438965	2
hacking	-0.11483341455459595	10.48360824584961	2
Hardcore	1.6741069555282593	9.003082275390625	5
Montgomery	6.803646087646484	6.262277126312256	3
unprecedented	-1.7028101682662964	9.924141883850098	2
harassment	-0.5528707504272461	10.79327392578125	2
posing	1.383348822593689	9.288904190063477	2
Forever	2.069767713546753	8.88358211517334	5
Cast	0.5098609328269958	7.945940971374512	5
genres	0.9534679651260376	7.238048553466797	1
automobile	2.512645721435547	7.692336559295654	6
vampires	0.5884708762168884	10.190460205078125	1
plague	-0.7767245769500732	10.169584274291992	2
Danielle	6.97674560546875	8.396866798400879	3
playlist	1.4596668481826782	6.024613857269287	4
AAA	3.2162654399871826	8.826193809509277	6
drummer	1.9723927974700928	5.3431782722473145	6
lure	0.11268114298582077	9.759452819824219	2
risky	-1.555166244506836	10.134655952453613	2
horny	1.541230320930481	10.35811996459961	0
hackers	-0.033982083201408386	10.514226913452148	6
cigarette	0.8945401310920715	11.94335651397705	6
Sterling	6.827090740203857	6.354888916015625	3
Angeles	5.436961650848389	7.543389797210693	5
60s	-2.0016801357269287	5.369662284851074	6
Drug	0.9586228728294373	11.808289527893066	5
Jewelry	3.0326333045959473	7.880982875823975	5
anal	1.4761115312576294	10.381134033203125	0
Youtube	1.2453817129135132	6.397546768188477	4
Murder	-0.16095872223377228	10.946276664733887	1
fountain	2.27164888381958	9.655920028686523	6
Jess	7.134767055511475	8.21857738494873	3
dragged	-0.21647103130817413	9.446146965026855	2
explosive	-1.5313316583633423	9.963665962219238	2
guitarist	1.964406132698059	5.4550981521606445	6
Devon	6.178282260894775	7.424752712249756	3
Tesla	2.5674455165863037	7.560958385467529	5
topping	-0.0005388688878156245	8.916844367980957	2
Writers	-0.12645933032035828	6.704282283782959	1
heroine	0.40497249364852905	7.2617411613464355	1
sophomore	3.0110092163085938	9.091180801391602	6
flashing	0.7544942498207092	9.308714866638184	2
majors	3.061494827270508	9.013519287109375	6
cocaine	1.0389877557754517	11.959938049316406	6
String	2.5373897552490234	6.564846992492676	5
adolescents	2.5172016620635986	9.698974609375	6
urgency	-1.3636952638626099	9.623807907104492	2
Ex	2.0875742435455322	8.641716003417969	5
Virgin	4.560483932495117	8.072890281677246	5
Passion	1.8520978689193726	8.89113712310791	5
virgin	1.4810357093811035	10.192098617553711	0
Manila	5.439361095428467	7.315455436706543	5
Extreme	1.6882688999176025	8.736119270324707	5
expedition	-1.2965612411499023	8.357213973999023	2
Ellis	7.109838962554932	5.814216136932373	3
teammates	3.06143856048584	8.774104118347168	6
Sonic	3.024073600769043	7.675568580627441	5
Alcohol	0.7972043752670288	11.900217056274414	5
borrowing	-0.6921449899673462	9.9021635055542	2
teenage	2.4399588108062744	9.75014591217041	6
Kristen	7.041386127471924	8.058907508850098	3
chicks	1.7808679342269897	10.423420906066895	0
drift	-0.5031179785728455	8.75005054473877	2
academy	2.946331262588501	8.944500923156738	6
exploded	-0.614287257194519	9.181503295898438	2
rebellion	-0.6293965578079224	10.511390686035156	2
Device	2.556791067123413	7.35893440246582	5
furious	-1.0793739557266235	10.021014213562012	2
Beatles	1.7798874378204346	5.665966987609863	6
Monroe	6.811456203460693	6.243123531341553	3
ringtone	1.3382309675216675	5.96427059173584	4
wreck	-0.7082769870758057	9.2241849899292	2
Swift	7.17802619934082	6.185894012451172	3
soaked	-0.33589646220207214	9.727310180664062	2
Kurt	7.4182868003845215	7.57134485244751	3
spike	-0.7524105310440063	8.34706974029541	2
obsession	0.28521254658699036	10.49612045288086	2
fuse	2.456005811691284	6.411447525024414	6
Powell	7.185937404632568	5.831844806671143	3
Straight	1.1035634279251099	8.777647972106934	5
1920s	-1.9134341478347778	5.483844757080078	6
Opinion	-0.5115939378738403	7.033322811126709	5
Bulgaria	5.48875617980957	6.673380374908447	5
multimedia	1.4096653461456299	6.291135787963867	4
Essex	5.735483169555664	7.281677722930908	5
AU	2.010260820388794	7.994429111480713	5
Kiss	1.9999487400054932	9.150120735168457	5
tense	-1.2265549898147583	9.778136253356934	2
thriller	0.34745651483535767	7.534586429595947	1
flawless	0.27634888887405396	9.445539474487305	2
filmmakers	0.4161917567253113	7.552913188934326	1
Gateway	3.3717141151428223	7.622344493865967	5
Released	-0.7505709528923035	7.042778968811035	5
plugs	2.43532657623291	6.350383281707764	6
rash	-1.0884393453598022	10.2152681350708	2
plea	-1.0230411291122437	10.802700996398926	2
broadcasting	1.504452109336853	6.398807048797607	4
Vanessa	6.820361137390137	8.482815742492676	3
Wave	2.0611608028411865	8.228014945983887	5
getaway	1.376395583152771	8.81847858428955	6
flooded	-0.47180110216140747	9.473139762878418	2
startups	-0.647438108921051	7.519559860229492	6
flick	0.45438939332962036	7.584543228149414	1
messaging	0.76160728931427	6.6683735847473145	4
Comics	0.9077627658843994	7.691514492034912	1
Shot	0.5047903060913086	8.366520881652832	5
tits	1.7559492588043213	10.554525375366211	0
Clarke	7.091495513916016	5.7199835777282715	3
Jared	7.534438133239746	7.436997890472412	3
Hilton	6.625476360321045	5.8398213386535645	3
Nikki	6.781497955322266	8.703042984008789	3
thieves	-0.23004773259162903	11.009952545166016	6
Identify	-1.1210938692092896	7.495390892028809	5
Courtney	7.121894836425781	8.067720413208008	3
Sebastian	6.96645450592041	7.580298900604248	3
robbery	-0.3014501929283142	11.102805137634277	2
HS	3.2378830909729004	8.906453132629395	5
poles	2.5641520023345947	6.328864574432373	6
Breaking	0.35328879952430725	8.56112289428711	5
texting	0.7849783897399902	6.557680606842041	4
tackles	-1.239597201347351	8.010701179504395	2
Scene	0.736935019493103	8.040545463562012	5
Leah	7.213194847106934	8.305093765258789	3
highs	-0.7979910969734192	8.235764503479004	2
Simpson	6.80922794342041	5.826579570770264	3
friction	-1.1734243631362915	9.25985050201416	2
shattered	-0.2952885925769806	9.31877613067627	2
Sierra	6.491036415100098	8.17175579071045	3
Whitney	7.07181978225708	8.212274551391602	3
mentors	2.797166585922241	9.392229080200195	6
Electronics	2.755739688873291	7.417089939117432	5
boiling	-0.4898918569087982	9.791828155517578	2
Rodriguez	6.3802266120910645	7.248143672943115	3
arrests	-0.5107215642929077	11.241975784301758	2
goddess	1.2304402589797974	10.23759651184082	0
discovers	-1.4389960765838623	7.940247058868408	2
secluded	0.9619359374046326	9.29511833190918	2
Budapest	5.556339740753174	6.712235450744629	5
Drugs	0.9471466541290283	11.843339920043945	5
outbreak	-0.8727861046791077	10.233213424682617	2
Titan	3.4066479206085205	8.058012008666992	5
Format	-0.2899511754512787	6.767615795135498	5
Trevor	7.5150885581970215	7.311032772064209	3
uploading	0.8665391206741333	6.199010848999023	4
formatting	0.09385678172111511	6.536550998687744	4
staging	-0.5193617343902588	10.514135360717773	2
revisions	0.11464520543813705	6.538741111755371	2
MTV	1.7466707229614258	6.437058448791504	4
Generation	2.0956099033355713	8.468546867370605	5
counselor	2.6732044219970703	9.512466430664062	6
Kat	6.977445602416992	8.652393341064453	3
Guitar	2.2728793621063232	5.678838729858398	5
Bedroom	2.328193187713623	8.564387321472168	5
Meyer	6.903242588043213	5.7874755859375	3
popping	0.12541663646697998	9.101627349853516	2
edits	0.0858146920800209	6.585296630859375	4
Miranda	6.7131266593933105	8.37136173248291	3
sparked	-1.804581642150879	8.532283782958984	2
Crisis	1.0838005542755127	8.090100288391113	5
Laurie	7.131009101867676	8.629291534423828	3
Garage	2.706843852996826	8.114779472351074	5
Brittany	6.95220422744751	8.293208122253418	3
Matrix	3.353231191635132	7.907247066497803	5
Pandora	1.761979341506958	6.058900833129883	4
Ivy	6.34832239151001	8.5925874710083	3
sa	1.753221035003662	7.6179280281066895	6
Gina	6.9525146484375	8.727588653564453	3
heroin	1.055137276649475	11.894195556640625	6
rebel	-0.48008421063423157	10.473091125488281	6
detention	-0.6013961434364319	11.283185005187988	2
XD	1.0266424417495728	7.572652339935303	1
nightlife	1.5452135801315308	8.73139476776123	6
trans	1.0243594646453857	10.57422924041748	6
Titans	3.3807852268218994	8.310944557189941	5
shotgun	0.28742823004722595	8.442305564880371	6
Clare	7.077027797698975	8.215275764465332	3
Fighting	0.6685670018196106	8.740058898925781	5
juices	1.391748309135437	10.810697555541992	0
Randall	7.371313571929932	7.080829620361328	3
Juliet	6.802319526672363	8.472638130187988	3
identities	0.9931595921516418	7.506978988647461	2
gigs	1.84736168384552	5.3524088859558105	6
urges	-1.2210673093795776	9.423742294311523	2
Jessie	7.083926200866699	8.144325256347656	3
sixteen	3.061544895172119	10.076240539550781	6
internship	-0.5214630961418152	6.452428817749023	6
Cardiff	5.762714385986328	7.176021575927734	5
Belfast	5.705542087554932	7.09772253036499	5
Tube	2.0747368335723877	6.901161193847656	5
Lightning	2.7482786178588867	7.7710747718811035	5
denim	2.0367815494537354	10.727599143981934	6
smash	-0.07470808178186417	9.327921867370605	2
crossover	0.9927706718444824	7.094207286834717	6
rocked	-0.3265468180179596	9.26388168334961	2
Minor	3.012888193130493	8.944398880004883	5
Horror	0.9044321775436401	7.7509331703186035	1
Valencia	5.966477870941162	7.422962665557861	5
implant	1.4409884214401245	11.041759490966797	6
dismissal	-0.8923922181129456	10.99571704864502	2
webcam	1.5633811950683594	10.217498779296875	4
modelling	1.4209513664245605	9.230147361755371	6
T-Mobile	2.986741304397583	7.290786266326904	4
Violence	-0.038886066526174545	10.865499496459961	1
Southeast	4.981378555297852	7.29585075378418	5
rehab	1.060604214668274	11.746644020080566	6
penetration	0.427021324634552	9.85054874420166	2
experimentation	-1.5099128484725952	8.183106422424316	2
‎	1.6807843446731567	7.826686382293701	5
Jade	6.2653679847717285	8.550535202026367	3
thirteen	3.070060968399048	10.080472946166992	6
Neal	7.263978958129883	6.26576566696167	3
crashing	-0.5116706490516663	9.25499153137207	2
Morrison	7.028788089752197	5.73899507522583	3
boobs	1.779117465019226	10.59243106842041	0
Alert	1.3978731632232666	7.6076860427856445	5
addicted	0.7848299741744995	11.512739181518555	6
asshole	1.5182008743286133	10.49505615234375	0
Dee	6.963582515716553	8.696965217590332	3
flashes	0.7600269913673401	9.246509552001953	2
desperation	-1.3430148363113403	9.651307106018066	2
Alexis	6.667655944824219	8.335143089294434	3
pearl	1.8562536239624023	9.610723495483398	6
hip-hop	1.6971728801727295	6.510363578796387	6
mag	-0.09553723782300949	7.8070526123046875	6
Rising	0.7784093618392944	8.624045372009277	5
lows	-0.773196280002594	8.24516773223877	2
Interviews	-0.31543558835983276	6.7698655128479	5
focuses	-1.2222068309783936	7.834629058837891	2
Riverside	5.573129177093506	7.471432209014893	5
amplifier	2.165585994720459	5.9607930183410645	6
Roses	2.2356488704681396	8.984395980834961	5
SM	3.862799644470215	7.769908905029297	5
Regina	5.878655433654785	7.9955058097839355	3
Electrical	2.7005457878112793	7.285609245300293	5
cannabis	1.1005308628082275	12.170788764953613	6
nerd	1.184746265411377	9.976090431213379	6
YA	0.786931574344635	7.331872463226318	1
Meg	7.237651824951172	8.409942626953125	3
vodka	0.9160598516464233	11.876473426818848	6
Gym	2.9123716354370117	8.421199798583984	5
Buffy	5.238526344299316	8.570269584655762	1
Elliott	6.984061241149902	5.9646453857421875	3
Winnipeg	5.343307971954346	7.610890865325928	5
Claudia	6.693763256072998	8.762384414672852	3
Halo	3.670172691345215	8.084696769714355	5
Warning	1.197979211807251	7.588276386260986	5
smashed	-0.2475067526102066	9.36684799194336	2
pounding	-0.19072145223617554	9.36972713470459	2
Kashmir	5.0515217781066895	6.7752580642700195	5
ace	3.207838773727417	9.094402313232422	6
Wire	2.6231324672698975	6.989317893981934	5
Darren	7.419701099395752	7.405276775360107	3
Rihanna	1.8240386247634888	6.2499799728393555	0
tunnels	2.118668794631958	6.910016059875488	2
shipments	0.9639450311660767	11.791245460510254	2
Covers	-0.6966230869293213	7.458865165710449	5
atheist	0.9377025961875916	10.53579044342041	0
Jenna	6.806153297424316	8.266623497009277	3
motel	2.3271236419677734	7.88209867477417	6
Erica	6.8627848625183105	8.582945823669434	3
kissed	1.5444386005401611	9.75068473815918	0
provocative	-1.6068705320358276	10.125691413879395	2
Romeo	6.602879047393799	8.075787544250488	3
strands	2.469191312789917	6.404903888702393	6
slut	1.4644511938095093	10.398792266845703	0
Hits	0.5045453906059265	8.367193222045898	5
booze	0.8540832996368408	11.971369743347168	6
whore	1.4799479246139526	10.472306251525879	0
Levi	7.534819602966309	7.694491863250732	3
racks	2.608222723007202	6.2072672843933105	6
Hayes	7.176786422729492	6.022608280181885	3
hail	-0.10356481373310089	8.989280700683594	2
fourteen	3.044215679168701	10.076144218444824	6
glamour	1.1127671003341675	9.114761352539062	6
au	1.7985323667526245	7.547237873077393	6
Bloom	6.851956367492676	5.731991767883301	3
spikes	-0.6507212519645691	8.310405731201172	2
Valerie	6.8500566482543945	8.800127029418945	3
Identity	1.4095467329025269	8.001129150390625	5
Zimmerman	7.041367053985596	6.203527927398682	3
subs	2.226144552230835	5.98792839050293	6
thighs	1.856049656867981	10.649487495422363	0
songwriter	1.832330346107483	5.58375358581543	6
Dakota	4.857434272766113	7.224747657775879	3
Renee	7.010674953460693	8.702736854553223	3
Laser	2.7737913131713867	7.448549747467041	5
panties	1.9721407890319824	10.669633865356445	0
Bahrain	5.624267101287842	6.6243767738342285	5
kidnapping	-0.4335609972476959	11.24803352355957	1
CW	2.191462755203247	6.78141450881958	5
start-up	-0.9240648746490479	7.506721019744873	6
homosexual	1.0119338035583496	10.551996231079102	0
groundbreaking	-1.7431894540786743	9.967143058776855	2
avenue	3.0429186820983887	7.142876625061035	2
Alexandra	6.680541515350342	8.58787727355957	3
Willow	6.086446285247803	8.520307540893555	3
Killer	1.1793735027313232	8.625629425048828	5
Shooting	0.48267248272895813	8.300670623779297	5
Harley	7.134190082550049	7.175386428833008	3
Nationals	2.799870491027832	8.653858184814453	5
convictions	-0.7156658172607422	11.199850082397461	2
Jasmine	6.3930888175964355	8.540717124938965	3
rapper	1.7383321523666382	6.151598930358887	6
Arrow	2.720423698425293	6.988824367523193	5
penetrate	0.19418573379516602	9.603241920471191	2
Initial	-1.2935739755630493	7.199334144592285	5
turmoil	-1.472595453262329	9.371620178222656	2
intrigue	0.2573138475418091	7.906332492828369	2
hash	1.006969690322876	12.110953330993652	6
recruits	3.120055675506592	9.009678840637207	6
graffiti	1.7583321332931519	9.551432609558105	6
pod	1.906937599182129	7.257319450378418	6
Oslo	5.504701614379883	6.853611946105957	5
Richards	7.153606414794922	5.870512008666992	3
drowning	-0.6802957057952881	9.630685806274414	2
injected	0.23017233610153198	10.087373733520508	2
offseason	3.1285011768341064	8.621307373046875	6
steroids	1.0868251323699951	11.676281929016113	6
drilled	0.4347888231277466	9.477062225341797	2
slash	-0.005486649926751852	9.278775215148926	2
paranoid	-1.320080041885376	10.021973609924316	2
Lindsey	7.14407205581665	8.073598861694336	3
Veronica	6.620692253112793	8.89877700805664	3
amps	2.268831491470337	5.972543716430664	6
DUI	0.1632499247789383	11.631085395812988	6
Savage	7.092886447906494	5.862707614898682	3
activism	-0.6802455186843872	10.648711204528809	6
welding	2.607006788253784	7.18979024887085	6
Automotive	2.741136074066162	7.6360955238342285	5
volatile	-1.530197262763977	9.964192390441895	2
slammed	-0.14102929830551147	9.367313385009766	2
Blonde	1.9777790307998657	9.284708976745605	0
Simmons	7.116506576538086	5.7227959632873535	3
Ricky	7.552036762237549	7.395429611206055	3
vain	-1.7147310972213745	9.667373657226562	2
coup	-0.6079126596450806	10.58339786529541	2
Anime	0.9571093916893005	7.640108585357666	1
reckless	-1.4584373235702515	10.200823783874512	2
Amateur	2.1514382362365723	9.310229301452637	5
Luis	6.457746505737305	7.785366535186768	3
nailed	0.3189798593521118	9.464950561523438	2
Warsaw	5.530288219451904	6.745269298553467	5
Corey	7.564338207244873	7.265881538391113	3
Mustang	2.823106050491333	7.826454162597656	5
}\fteneffectsizevis

\DeclareCaptionStyle{ruled}{labelfont=normalfont,labelsep=colon,strut=off} 
\floatstyle{ruled}
\newfloat{listing}{tb}{lst}{}
\floatname{listing}{Listing}

\title{Representation Bias of Adolescents in AI: A Bilingual, Bicultural Study}

\author {
    Robert Wolfe\textsuperscript{\rm *},
    Aayushi Dangol\textsuperscript{\rm *},
    Bill Howe,
    Alexis Hiniker
}
\affiliations {
    University of Washington \\
    \textsuperscript{\rm *}Equal Contribution\\
    rwolfe3@uw.edu, adango@uw.edu, billhowe@uw.edu, alexisr@uw.edu
}

\begin{document}

\maketitle

\begin{abstract}
  Popular and news media often portray teenagers with sensationalism, as both \textit{a risk} to society and \textit{at risk} from society. As AI begins to absorb some of the epistemic functions of traditional media, we study how teenagers in two countries speaking two languages: 1) are depicted by AI, and 2) how they would prefer to be depicted. Specifically, we study the biases about teenagers learned by static word embeddings (SWEs) and generative language models (GLMs), comparing these with the perspectives of adolescents living in the U.S. and Nepal. We find English-language SWEs associate teenagers with societal problems, and more than 50\% of the 1,000 words most associated with teenagers in the pretrained GloVe SWE reflect such problems. Given prompts about teenagers, 30\% of outputs from GPT2-XL and 29\% from LLaMA-2-7B GLMs discuss societal problems, most commonly violence, but also drug use, mental illness, and sexual taboo. Nepali models, while not free of such associations, are less dominated by social problems. Data from workshops with N=$13$ U.S. adolescents and N=$18$ Nepalese adolescents show that AI presentations are disconnected from teenage life, which revolves around activities like school and friendship. Participant ratings of how well 20 trait words describe teens are decorrelated from SWE associations, with Pearson's $\rho$=.02, \textit{n.s.} in English FastText and $\rho$=.06, \textit{n.s.} in GloVe; and $\rho$=.06, \textit{n.s.} in Nepali FastText and $\rho$=$-$.23, \textit{n.s.} in GloVe. U.S. participants suggested AI could fairly present teens by highlighting \textit{diversity}, while Nepalese participants centered \textit{positivity}. Participants were optimistic that, if it learned from \textit{adolescents}, rather than media sources, AI could help mitigate stereotypes. Our work offers an understanding of the ways SWEs and GLMs misrepresent a developmentally vulnerable group and provides a template for less sensationalized characterization.
\end{abstract}
\section{Introduction}

\begin{figure*}[!htbp]
\centering
\begin{tikzpicture}[baseline]
    \begin{axis} [
        height=4.81cm,
        width=.48\textwidth,
        ybar = .05cm,
        bar width = 16.5pt,
        ymin = 0, 
        ymax = 40,
        ylabel=\% of Prompt Continuations,
        ylabel style={font=\footnotesize},
        ylabel shift=-2pt,
        xtick = {1,2,3,4,5},
        xtick style={draw=none},
        xlabel={\footnotesize English \ \ \ \ \ \ \ \ \ \ \ \ \ \ \ \ \ \ \ \ Nepali},
        ytick pos = left,
        bar shift=0pt,
        ymajorgrids=true,
        xticklabels = {LLaMA-2, GPT-2 EN, Teen EN, GPT-2 NE, Teen NE},
        title= Teenager Prompt Continuations Describing Social Problems,
        xticklabel style={font=\scriptsize},
        enlarge x limits={abs=1cm},
        legend style={at={(0.52,0.72)},
        anchor=south west, 
        legend columns=1,
        font=\footnotesize},
        title style={font=\footnotesize},
    ]    
    \addplot [pattern=crosshatch, pattern color = red] coordinates {(1,29) (2,30) (4,13) (3,0) (5,0)};    
    \addplot [pattern=vertical lines, pattern color = blue] coordinates {(1,0) (2,0) (4,0) (3,4) (5,1)};    
    \legend {Generative Model, Teen Participants};
    \end{axis}
    \end{tikzpicture} \hspace{.8cm}
    \begin{tikzpicture}[baseline]
    \centering
    \begin{axis}[
    legend pos=outer north east,
    height=4.81cm,
    width=.48\textwidth,
    ymajorgrids=true,
    ylabel={\footnotesize UMAP Dim 2},
    xlabel={\footnotesize UMAP Dim 1},
    xtick style={draw=none},
    xticklabel style={font=\scriptsize},
    yticklabel style={font=\scriptsize},
    title={\footnotesize FastText-EN Words Uniquely Associated with Teenagers},
    ylabel near ticks,
        ylabel shift=-2pt,
    legend style={
        at={(.78,.58)}, 
        anchor=south,
        font=\scriptsize,
        legend columns = 2
        }
    ]
    
    \addplot[scatter,
        only marks,
        point meta=explicit symbolic,
        visualization depends on=\thisrow{cluster} \as \class,
        scatter/classes={
            0={mark=*,mark size=2.5pt,color=red!80,draw=black, opacity=.75},
            1={mark=*,mark size=2.5pt,color=green!80,draw=black, opacity=.75},
            2={mark=*,mark size=2.5pt,color=blue!80,draw=black, opacity=.75},
            3={mark=*,mark size=2.5pt,color=brown!80,draw=black, opacity=.75},
            4={mark=*,mark size=2.5pt,color=orange!80,draw=black, opacity=.75},
            5={mark=*,mark size=2.5pt,color=purple!80,draw=black, opacity=.75},
            6={mark=*,mark size=2.5pt,color=yellow!80,draw=black, opacity=.75}%
        }
    ]
    table[x=dim1, y=dim2, meta=cluster] 
    {\fteneffectsizevis};

    \draw[-] (-.5,11.3) -- (-1.3,11.4) node[left] {\scriptsize arrested};

    \draw[-] (1.67,10.31) -- (2.0,11.4) node[above] {\scriptsize nude};

    \draw[-] (.85, 11.97) -- (.23,12.15) node[left] {\scriptsize booze};

    \draw[-] (.5, 6.63) -- (0,5.7) node[below] {\scriptsize texts};

    \addlegendentry{Sex}
    \addlegendentry{Teen Media}
    \addlegendentry{Violence}
    \addlegendentry{Names}
    \addlegendentry{Tech}
    \addlegendentry{Places}
    \addlegendentry{Rebellion}
    
    \end{axis}
    \end{tikzpicture}
    
    \caption{\footnotesize Left: Teenage participants were much less likely to continue prompts about teenagers with social problems than were GLMs. Right: Words associated with adolescents over any other age group in English SWEs reflect violence, rebellion, and sexualization.}
    \label{fig:teaser}
\end{figure*}

Teenagers feature more prominently in western media accounts of new technologies than perhaps any other user group. They are the group most likely to adopt and capably use new technologies, including social media \cite{pew2023socialmedia} and ChatGPT \cite{hill2023chatgpt}. However, to read media accounts, they are also the most likely to \textit{misuse} new technologies, leading to harm to others, or inadvertent harm to themselves \cite{stern2017constructing}. Such narratives have consequences for adolescent access to technology: concerns about compulsive use of social media, cyberbullying, and sexual predation led to a March 2024 ban on use of numerous social media platforms by younger teenagers in the state of Florida \cite{guardian2024desantis}. Concerns about deceptive design and online safety warrant consideration; yet the response---a blanket ban---suggests a framing that emphasizes the danger of adolescent technology use and affords adolescents little agency.

Such presentations continue a decades-long trend in western media portraying teenagers as simultaneously \textit{a risk} to society and \textit{at risk} from society \cite{pain2003youth}. Though largely disconnected from most adults' experiences with teens \cite{aubrun2000aliens}, media portrayals of adolescents have centered violence, drug abuse, hyper-sexualization, technology addiction, and even religious fanaticism as pressing issues that warrant responses ranging from targeted media campaigns to government legislation \cite{clark2005angels,marwick2008catch,glassner2010culture,telzer2022challenging}. Though such portrayals appear sensationalistic in hindsight, representations of teenagers in media sources nonetheless shape adults' beliefs about what adolescents are like, influencing the treatment of adolescents in public places \cite{bernier2011representations} and the restrictiveness of policy intended to influence adolescent behavior \cite{dorfman2001off}, thus directly affecting how teens are treated in practice.

In the present work, we study societal attitudes toward adolescents learned by static word embeddings (SWEs) and generative language models (GLMs), comparing with attitudes reported by adolescents themselves. Because prior work suggests attitudes toward adolescents vary across cultures \cite{larson2004adolescence,di2023predictors}, we undertake a bilingual, bicultural study, examining U.S. attitudes and English-language models, as well as models trained on Nepali, a low-resource language spoken primarily in Nepal, a South Asian country in the Global South, and a native language for a first author of this work. We held workshops with $N$=13 English-speaking adolescents in the U.S. and $N$=18 Nepali-speaking adolescents in Nepal, asking how adolescents \textit{are} represented in media, and how they \textit{should} be represented in AI. We make three contributions: 

\begin{itemize}
    \item \textbf{We show that English-language SWEs and GLMs associate adolescents predominantly with social problems.} Clustering the 1,000 words most associated with teenagers in English GloVe and FastText SWEs reveals that clusters related to drugs, rebellion, violence, mental illness, stereotypes, and sexual taboo account for more than 50\% of words in GloVe and more than 40\% in FastText. Similarly, using prompts about teenagers derived from \citet{stern2005self}, we show that 29\% of English LLaMA-2-7B outputs and 30\% of GPT2-XL outputs depict societal problems. Of these, 47\% depict violence in LLaMA-2, and 50\% depict violence in GPT2-XL. Many such outputs mimic the format of ``high-quality'' training data---newspapers and journalistic media. Only 13\% of distilGPT2 Nepali continuations reflect societal problems, and 10.1\% of the words most associated with teenager in Nepali GloVe describe societal problems.
        
    \item \textbf{We show that AI representations are disconnected from adolescent self-perceptions.} Adolescent ratings of their own traits are decorrelated from SWE associations between corresponding trait vectors and the \textit{teenager} vector, with Pearson's $\rho$=.02, \textit{n.s.} in FastText and $\rho$=.06, \textit{n.s.} in GloVe for English; and $\rho$=.06, \textit{n.s.} in FastText and $\rho$=$-$.23, \textit{n.s.} in GloVe for Nepali. Participant continuations of the same prompts used with GLMs show social problems arise in fewer than 4\% of U.S. participant continuations and fewer than 1\% of Nepalese participant continuations (Fig \ref{fig:teaser}).
    
    \item \textbf{We discuss two central concerns of participants for fair representation in AI: diversity and positivity.} U.S. and Nepalese participants were aware of adolescent media stereotypes, and noted the difficulty in achieving fair representation. U.S. participants stressed that AI should foreground the \textit{diversity} of teenagers, while Nepalese participants stressed that AI should present the positive traits of teenagers. Both groups expressed optimism that AI could correct media stereotypes about adolescents.
\end{itemize}

\noindent Our work shows that GLMs learn societal biases latent in media framings. As user-facing GLMs are integrated into schools and other contexts where they will impact adolescents' lives, research must center participatory approaches to AI \cite{delgado2023participatory} to ensure groups with less agency, like adolescents, are represented in ways that capture not a media presentation but a group's understanding of itself.
\section{Related Work}

We review prior work on depictions of adolescents in popular and news media, sources often used to train AI. We then consider the language models studied and age biases in AI.

\subsection{Defining Adolescence}

The National Institutes of Health (NIH)  define Adolescents as persons between 13 and 17 years old, distinct from Children (1 through 12), Adults (18 and older), and Older Adults (65 and older) \cite{NIHages}. While definitions may vary between cultures and across time \cite{arnett1999adolescent}, we adopt the NIH definition, which is consistent with related work.

\subsection{Media Representations of Adolescents}

Prior work finds that popular and news media depictions of adolescents are generally negative, with positive interactions involving teenagers portrayed as deviations from the norm \cite{bernier2011representations}. News coverage of teenagers often depicts supposed epidemics of violence, crime, drug abuse, mental illness, and immorality, which are usually not well supported by evidence \cite{glassner2010culture,telzer2022challenging}. In foundational work, \citet{dorfman1997youth} find that most California TV news reports related to violence feature youth, and that only education policy receives as much treatment as violence in newspaper coverage about adolescents. \citet{males1999framing} find that LA Times articles included adolescents in stories about violence five times more frequently than adults. Adolescent behavior may be presented as dangerous even when not volitional, as \citet{best2008teen} find that activities as simple as teenage driving can be framed as pressing issues in the media. More recently, teenage use of technology has become a subject of public concern, and \citet{stern2017constructing} find that print and online news media portray teens as having an unhealthy relationship with social media. Previously, \citet{stern2005self} found that U.S. films depict teenagers as violent, self-absorbed, and disengaged from civic life. As discussed in the Methods, we draw on \citet{stern2005self} to create GLM prompts.

\subsubsection{Societal Impact}

Media depictions shape adult views of adolescents and may shape adolescent behavior. \citet{hancock2001school} shows that adults overestimate and perceive illusory increases in adolescent crime. \citet{aubrun2000aliens} find most adults report good experiences with teenagers they know but consider such experiences atypical, rather than questioning media framing. \citet{dorfman2001off} note that negative media portrayals, especially of adolescents of color, lend justification to harsher treatment and more restrictive policies. Moreover, \citet{qu2020early} find that younger teens' \textit{own} beliefs in teenage stereotypes contribute to behavioral problems. \citet{buchanan2023adolescent} argue that, to prevent a self-fulfilling prophecy, descriptions of adolescent ``stress and storm'' must be replaced with a less reductive framing, such as ``possibility and promise.''

\subsubsection{Societal Variation}

Though some aspects of adolescence appear consistent around the world \cite{steinberg2018around}, scholars describe significant variation in characterizations of adolescence both within and across cultures \cite{buchanan2023typicality}. \citet{enright1987economic} note that definitions of adolescence change over time based on society’s needs: during war time, teens are portrayed as rugged and adultlike, but when not desired in the workforce, teens are portrayed as more childlike. \citet{arnett1999adolescent} note that adolescent stress may be more pronounced in individualistic western cultures, while \citet{larson2004adolescence} use the plural form ``adolescences'' to describe variations around the world and across time, noting that teen years are not consistently characterized by emotional turmoil and psychic separation from parents. Finally, \citet{di2023predictors} observe differences in emotion regulation in teenagers in Italy and Colombia, suggesting cultural factors play a role in adolescent well-being.

\subsection{Language Models}

In this work, we study \textbf{static word embeddings} (SWEs) and \textbf{generative language models} (GLMs). SWEs are trained using deep neural networks (DNNs) to represent words as vectors based on the conditional probability of their co-occurrence with surrounding words \cite{mikolov2013distributed, collobert2011natural}. We study FastText \cite{bojanowski2017enriching}, an extension of Word2Vec \cite{mikolov2013linguistic} that incorporates subword information, and Global Vectors for Word Representation (GloVe) \cite{pennington2014glove}, which incorporates corpus-level statistics to improve semantics. SWEs are now widely used in social science \cite{bhatia2023predicting,guan2024have} to study societal attitudes \cite{garg2018word}, because the cosine distance between word vectors captures information about semantic similarity \cite{hill2015simlex}. GLMs are DNNs based on the transformer architecture \cite{vaswani2017attention} that learn to predict the next token (word or subword) \cite{radford2018improving}. GLMs allow users to interact with a model by ``prompting'' it---providing text input for continuation by the model \cite{brown2020language}. Models like ChatGPT \cite{openai2022chatgpt} fine-tune a pretrained GLM to follow user instructions and adhere to user preferences \cite{ouyang2022training}. We study GPT2 \cite{radford2019language}, the last GLM released publicly by OpenAI and the most-downloaded GLM in the Transformers library \cite{wolf-etal-2020-transformers}, and Meta's LLaMA-2 \cite{touvron2023llama2}, an open-weight model from which dozens of open-weight chatbots have been trained \cite{vicuna2023,alpaca,wolfe2024laboratory}. We avoid proprietary models like ChatGPT due to uncertain reproducibility of results from models for which weights are unavailable \cite{liesenfeld2023opening}.

\subsubsection{Low-Resource Languages} Nepali is a ``low-resource'' language, meaning that much less text data exists for training Nepali NLP models than other languages \cite{besacier2014automatic}, and model performance is likely to lag behind that of higher-resource languages such as English \cite{ranathunga2023neural}. While a multilingual model may improve performance in a low-resource language \cite{scao2022bloom}, its representations may also take on semantic properties and biases of a higher-resource languages (e.g., English) \cite{zhao-etal-2020-gender,ramesh2023fairness}. Thus, our work requires monolingual technologies to ensure we capture semantic properties of the intended language, rather than the semantic influence of a higher-resource language.

\subsection{Age Biases in AI}

Research on age biases in AI describes technical failures of technologies like emotion recognition for older adults \cite{kim2021age}, precipitated by underrepresentation in training data \cite{park2021understanding}. Studies of young/old bias in SWEs find that youth is preferable to old age \cite{caliskan2017semantics,diaz2018addressing,swinger2019biases} but do not analyze adolescents as a distinct age group. Most similar to our work are studies of biases in multimodal language-vision models. \citet{agarwal2021evaluating} find that OpenAI's CLIP \cite{radford2021learning}  associates criminality with images of adolescents, while \citet{wolfe2023contrastive} find text-to-image generators like Stable Diffusion \cite{rombach2022high} output sexually objectifying images of teenage girls.
\section{Models and Training Data}

The present work studies monolingual SWEs and GLMs in English and in Nepali. We examine the following \textbf{SWEs}:

\begin{itemize}
    \item \textbf{GloVe-CC}, 300-dimensional (300d) English-language GloVe embeddings pretrained by \citet{pennington2014glove} on the 840-billion token Common Crawl circa 2014 \cite{CommonCrawl}.
    \item \textbf{FastText-CC}, 300d FastText embeddings pretrained by \citet{bojanowski2017enriching} on a filtered and deduplicated version of Common Crawl.
    \item \textbf{GloVe-NE}, 300d GloVe embeddings trained by the authors, discussed further below.
    \item \textbf{FastText-NE}, 300d FastText embeddings pretrained by \citet{grave2018learning} on Nepali Wikipedia.
\end{itemize}

\noindent FastText embeddings like FastText-NE are among the most used low-resource models for social science \cite{lindqvist2022most}. We trained a Nepali GloVe embedding after considering several pretrained Nepali embeddings, including the NPVec1 model of \citet{koirala-niraula-2021-npvec1}, the Nepali Word2Vec model of \citet{dz6s-my90-19}, and the model of \citet{10.1145/3582768.3582799}. We ultimately trained an embedding on the dataset of \citet{timilsina2022nepberta} because it contained three times the data (800 million tokens from 2.76 million Nepali webpages) as used to train any other model, allowing us to produce an embedding more comparable in scale to English-language GloVe. Our training hyperparameters adhered closely to best practices for GloVe.

We also study the following pretrained \textbf{GLMs}:

\begin{itemize}
    \item \textbf{OpenAI GPT2-XL}, an English-language GLM trained on OpenAI's WebText dataset \cite{radford2019language}.
    \item \textbf{Meta LLaMA-2-7B}, an English-language GLM trained on public datasets including The Pile \cite{gao2020pile}.
    \item \textbf{DistilGPT2 Nepali}, an open-weight, reduced-parameter version of GPT2 pretrained on the nepalitext dataset, which consists of Nepali text from the CC100 \cite{wenzek-etal-2020-ccnet} and OSCAR \cite{OrtizSuarezSagotRomary2019} datasets, as well as Nepali Wikipedia.
\end{itemize}

\noindent We use 4-bit quantization \cite{dettmers2024qlora} to mount LLaMA-2-7B on affordable GPU hardware. Our code is available at https://github.com/wolferobert3/adolescent-representation-bias.
\section{Methods}\label{sec:methods}

We use mixed quantitative and qualitative methods to collect and analyze the presentations of adolescence in AI and those reported by adolescent participants in our study.

\subsection{Computational Methods}

We obtained data from the SWEs and GLMs by employing methods appropriate to the models' pretraining objectives.

\subsubsection{SWEs}

For each SWE, we computed 1) the 1,000 words \textit{most} associated with adolescents; and 2) the 1,000 most frequently occurring words \textit{uniquely} associated with adolescents over any other age group. Given an embedding vocabulary $V$, we define an Adolescent target group $A$.

$A$, Teenager: \textit{teenager, teenagers, teen, teens, teenage, teenaged, adolescent, adolescence}

\noindent To obtain the \textit{most} associated words with $A$, We compute the mean cosine similarity $s=\frac{\sum_{a \in A}{cos(\vec{w}, \vec{a})}}{|A|}$ for every word vector $\vec{w}$ corresponding to a word $w \in V$, and select the words with the 1,000 largest values of $s$. 

To obtain the highest frequency words \textit{uniquely} associated with $A$, we use a Single-Category Word Embedding Association Test (SC-WEAT) \cite{caliskan2017semantics,caliskan2022gender} to compare the relative similarity of a word $w$ to two attribute groups $A$ and $B$: \looseness=-1

\begin{equation}
d(w,A,B) = \frac{\textrm{mean}_{a\in A}\textrm{cos}(\vec{w},\vec{a}) - \textrm{mean}_{b\in B}\textrm{cos}(\vec{w},\vec{b})}{\textrm{std\_dev}_{x \in A \cup B}\textrm{cos}(\vec{w},\vec{x})}
\end{equation}

\noindent The SC-WEAT returns an effect size (Cohen's $d$) and a $p$-value based on a permutation test. Unlike some SC-WEATs, which define $A$ and $B$ based on two poles of a binary (e.g., Male/Female), Teenager has no clear opposing pole for $B$. Thus, we define three $B$ groups using the age ranges specified by the NIH: Children ($B_1$), Adults ($B_2$), and Older Adults ($B_3$):

\begin{itemize}
    \item $B_1$, Children: \textit{child, children, childlike, childhood, kid, kids,  schoolchild, schoolchildren}
    \item $B_2$, Adult: \textit{adult, adults, adulthood, middle-age, middle-aged, grownup, grown-up, grownups}
    \item $B_3$, Older Adults: \textit{aged, aging, older, old-age, elder, elders, elderly, retiree}
\end{itemize}

\noindent For every word $w \in V$, an SC-WEAT is taken between $A$ and $B_1$; $A$ and $B_2$; and $A$ and $B_3$. We select only words that exhibit large, statistically significant effect sizes with $A$ when compared with all three of $B_1$, $B_2$, and $B_3$. Formally, let $W_1$ denote the words $w$ in $V$ such that $d(w,A,B_1) > 0.8, p<.05$; $W_2$ the words such that $d(w,A,B_2) > 0.8,  p<.05$; and $W_3$ the words such that, $d(w,A,B_3) > 0.8, p<.05$. That is: \looseness=-1

\begin{equation}
\centering
\begin{split}
    W_1 = \{w \in V \ | \ d(w,A,B_1) > 0.8, p<.05 \} \\
    W_2 = \{w \in V \ | \ d(w,A,B_2) > 0.8, p<.05 \} \\
    W_3 = \{w \in V \ | \ d(w,A,B_3) > 0.8, p<.05 \}
\end{split}
\end{equation}

\noindent We select the  words $W_{A}$ exhibiting a large, significant effect with Adolescent over Children, Adults, and Older Adults:

\begin{equation}
    W_{A} = W_1 \cap W_2 \cap W_3
\end{equation}

\noindent We then select from $W_{A}$ the 1,000 most frequently occurring words in the corpus that produced $V$, a straightforward task because SWEs are rank-ordered based on word frequency. Word groups were constructed by 1) referring NIH descriptions of each age range; and 2) using WordNet \cite{miller1995wordnet} to increase the number of words in each group to eight, meeting the SC-WEAT minimum \cite{caliskan2022gender}. The Nepali-speaking first author translated $A$, $B_1$, $B_2$, and $B_3$ into Nepali for use with the Nepali embeddings. We provide these translations in our code repository. 

\begin{table}[]
    \centering
    \fontsize{8.25}{8.75}\selectfont
    \begin{tabular}{|c|c|}
    \toprule
    Domain & Generative Language Model Prompts  \\
    \midrule
    \multirow{5}{*}{Behaviors} & At school, the teenager\dots \\
    & At home, the teenager\dots \\
    & At work, the teenager\dots \\
    & At the party, the teenager\dots \\
    & Online, the teenager\dots \\
    \hline
    \multirow{5}{*}{Motivations} & The teenager worked because they wanted\dots \\
    & The teenager studied because they wanted\dots \\
    & The teenager flirted because they wanted\dots \\
    & The teenager socialized because they wanted\dots \\
    & The teenager volunteered because they wanted\dots \\
    \hline
    \multirow{5}{*}{Relationships} & With their friends, the teenager\dots \\
    & With their parents, the teenager\dots \\
    & With their teachers, the teenager\dots \\
    & With their coworkers, the teenager\dots \\
    & With their romantic partner, the teenager\dots \\
    \bottomrule
    \end{tabular}
    \caption{\footnotesize Prompts for GLMs, drawing on the work of \citet{stern2005self}.}
    \label{tab:lm_prompts}
\end{table}

\subsubsection{GLMs} We study GLMs by using them to generate text conditioned on a prompt. Table \ref{tab:lm_prompts} includes the prompts we designed, drawing on the prior work of \citet{stern2005self}, who examined media portrayals of the behaviors, motivations, and relationships of adolescents. Prompts are designed to be 1) consistent with the GLM's pretraining objective; 2) non-leading and possible to answer in an unbiased manner; and 3) easily adaptable for the human subjects study described below. Prompts for the Nepali GLM are translations by the first author and provided in our code repository.

To generate text, we use multinomial sampling with the temperature set to 1.0, allowing the GLM to sample next words based on its probability distribution over the output vocabulary \cite{ippolito-etal-2019-comparison}. This allows us to generate 15 distinct continuations for each prompt (225 per model) that are high-probability for the GLM and representative of its semantic associations. GLMs are restricted to produce no more than 50 new tokens (words or subwords) of output.  \looseness=-1

\subsection{Workshop Sessions}

We held workshops on Zoom with $N$=14 English-speaking adolescents in the U.S. and $N$=18 Nepali-speaking adolescents in Nepal. Our university's IRB approved this study.

\subsubsection{Participants}

We used purposive sampling \cite{campbell2020purposive} to recruit two populations of participants: English-speaking adolescents between 13 and 17 residing in the United States, and Nepali-speaking adolescents between 13 and 17 residing in Nepal. To recruit U.S. participants, we used a contact list of parents who indicated their willingness to be contacted by our university regarding enrolling their children in research. We sent one email to individuals whose children met our inclusion criteria, then called them once at the phone number provided. To recruit Nepalese participants, a relative of the first author residing in Kathmandu posted recruiting flyers at two Kathmandu high schools. We collected signed assent forms from participants and signed consent forms from their parents. U.S. participants received \$25 Amazon credit. Because Amazon does not operate in Nepal (nor does any equivalent), we compensated participants in Nepal via direct payment equal to \$7.50 USD in Nepalese Rupees, after consulting a relative of the first author living in Nepal regarding exchange rate to ensure we did not bias participant responses \cite{millum2019payment}.

\subsubsection{Workshop} All workshops took place over Zoom during December 2023 and January 2024. Participants could choose a synchronous or asynchronous format. With exception of a session wherein two participants asked to join a workshop together, we conducted workshops individually to allow participants more opportunities to ask questions. Sessions began with a five minute, story-based introduction to how AI learns language---for example, by guessing the next word in a sentence, or arranging words based on their similarity to each other. Participants were then asked to help AI learn about teenagers, which involved the following tasks: \looseness=-1

\begin{itemize}
    \item Write the top ten words that come into your head when you hear the word \textit{teenager}.
    \item Write ten words that \textit{only} describe teenagers, and do not describe children, adults, or older adults.
    \item Complete the sentence with a few words, using the GLM prompts provided in Table \ref{tab:lm_prompts}.
    \item Rate 20 traits on a scale from 1 (most similar) to 5 (least similar) based on how well they describe teenagers.
    \item Provide the AI with instructions on how to discuss teenagers fairly (both accurately and without bias).
\end{itemize}

\noindent Participants were asked to write about whether and why AI should learn about teenagers from teenagers themselves, rather than media sources. Finally, we engaged in dialogue with synchronous participants to answer their questions about AI. Asynchronous participants watched a video recorded by the research team and were provided with the emails of the first two authors for any questions. U.S. participants completed the research instruments using a Google Form, while Nepalese participants completed research instruments using paper, and sent photos of the worksheets to the authors, who transcribed them for further analysis.

\subsection{Data Analysis}

We followed a Directed Content Analysis methodology \cite{assarroudi2018directed} to analyze data from models and participants. We first used k-means clustering on the word vectors most associated and uniquely associated with adolescents in the GloVe-CC, GloVe-NE, FastText-CC, and FastText-NE embeddings. We selected the number of clusters (between 5 and 10) using Silhouette Score \cite{rousseeuw1987silhouettes}. The first two authors then individually reviewed the clusters and assigned labels (e.g., a cluster containing \textit{Justin, Morgan, etc.}, was assigned \textit{First Names}). The authors then met to discuss and formalize labels into initial codes. The authors then applied the codes to the GLM outputs. Where an output did not belong to any existing code, it was added to an \textit{Other} category. After coding the output of each GLM, the authors met to review outputs classified as \textit{Other}, and decided whether to add new codes. The authors discussed output on which they did not agree and either resolved the code in discussion or added it to the \textit{Other} category if agreement was not reached. Multiple codes were applied to an output if appropriate.

Next, the authors applied the codes to participant workshop data, adding codes as needed and keeping track via memos of how participant responses differed from model outputs. The authors sequentially reviewed the word similarity, prompt continuation, and instructions for AI fairness data, meeting to discuss and resolve differences after each phase of coding. All data was coded in Google Sheets, and each author was provided with separate copies of model and participant data so that the authors could not see each other's codes before discussion. The Nepali-speaking first author translated Nepali content and provided guidance where the meaning of a translation was uncertain. After arriving at a final hierarchy of 40 codes with 10 top-level codes such as \textit{Teen Experiences} and \textit{Law and Crime}, the authors reviewed all model and human materials again, refining code assignments as appropriate.

The authors then met three times to arrive at themes describing the findings. During the first meeting, the authors used affinity diagramming to visualize proposed themes that were shared across languages and data sources (model or human) and those which were distinct across languages and sources. After this meeting, the authors wrote memos describing the proposed themes. The authors shared the memos and discussed them in the second meeting to arrive at the final themes. The authors then collected representative quotes and model output, which they reviewed in the third meeting, and prepared for inclusion in the Results.
\begin{table*}
\fontsize{6.25}{7}\selectfont
\centering
\begin{tabular}{c|ccc||ccc}
\toprule
 \multicolumn{4}{c||}{\textbf{\textit{Most}} Associated Words (English)} & \multicolumn{3}{c}{\textbf{\textit{Most}} Associated Words (Nepali)} \\
 \hline
\textbf{$E$} & \% & \textbf{Cluster Name} & \textbf{Representative Words} & \% & \textbf{Cluster Name} & \textbf{Representative Words} \\
\hline
\multirow{3}{*}{FT} & 14.7 & Teenagers & teenagers, teens, youths, juveniles, screenagers  & 9.0 & Teens (female) & young woman, girl, young girl, woman, girl child \\
& 12.4 & Teen Years & 19-year-old, fifteen-years-old, then-16-year-old & 8.2 & Teens (male) & adolescent, youthful, young people, boyhood, young man \\
& 9.5 & Other Ages & college-student, mid-twenties, baby-boomers & 1.1 & Age Groups & young, youth, of age, adult, child, elderly, very young \\
& 8.1 & School & high-schooler, middle-schooler, school-age, ninth-grade & 15.0 & Teen Names & Surkishore, Amar Kishore, Ranjeeta, Junita, Amritraj \\
& 6.6 & Puberty & puberty, pimples, gawkiness, growing-up, juvenility & 34.2 & Life Changes & puberty, menstruation, marriageable, employable, widow \\
& 10.0 & Coming of Age & coming-of-age, right-of-passage, prom-night, semi-adult & 27.6 & Relationships & lovers, marriage, friends, mother-son, siblings \\
& 8.8 & Stereotypes & acne-ridden, braces-wearing, sullen, spiky-haired & 4.8 & Cultural Figures & princess, divine girl, Sukanya, Dakshyakanya \\
& 9.7 & Rebellion & rebellious, sex-crazed, drug-crazed, angst-filled & & & \\
& 1.6 & Delinquency & delinquents, punks, runaways, juvey, gang-involved & & & \\
& 18.6 & Sex & barely-legal, underage, jail-bait, impressionable & & &  \\
\hline
\multirow{3}{*}{GloVe} & 11.2 & Age Words & 16-year-old, 14-year-old, youngster, prodigy  & 17.5 & Teenagers & young women, young girl, youth, junior, generations \\
& 8.5 & Relationships & dad, mom, friends, lover, teacher, classmates & 15.3 & Relationships & father, son, daughter, couple, brother \\
& 15.6 & Stereotypes & jocks, nerd, emo, punks, stoned, self-absorbed & 4.9 & School & school, class, students, principal, studious \\
& 12.0 & Mental Illness & self-esteem, psychotic, antisocial, suicidal & 21.7 & Names & Rana, Ashish, Lalit, Mohan, Uttam \\
& 11.8 & Risks & at-risk, dropout, homeless, pregnancies, inner-city & 11.8 & Times & morning, year, Baisakh (month), Magh (month) \\
& 18.6 & Violence & violent, bullied, victim, murder, suicide & 10.1 & Violence & fugitive, murder, kidnapped, police \\
& 13.1 & Sex & horny, masturbating, kinky, seduce, lusty & 18.6 & Public Events & demonstration, committee, program, district \\
& 9.2 & Sexual Taboo & taboo, underage, lolita, interracial, voyeur & & &  \\
\hline
 \multicolumn{4}{c||}{\textbf{\textit{Exclusively}} Associated Words (English)} & \multicolumn{3}{c}{\textbf{\textit{Exclusively}} Associated Words (Nepali)} \\
 \hline
\textbf{$E$} & \% & \textbf{Cluster Name} & \textbf{Representative Words} & \% & \textbf{Cluster Name} & \textbf{Representative Words} \\
\hline
\multirow{3}{*}{FT} & 16.0 & First Names & Sam, Justin, Morgan, Madison, Bailey & 9.8 & Internet & URL, portal, Fedora, Photos, Yahoo, interface \\
& 21.5 & Places \& Headlines & Seattle, London, Campus, Driver, Youth & 58.3 & Travel \& Tourism  & attractions, ambassador, architecture, places, Janakpur \\
& 5.8 & Teen Media & vampire, manga, YA, murder, zombies & 25.6 & Media \& Names  & BBC, FM, Youtube, Times magazine, Kishor, Pramod \\
& 5.5 & Technology & texts, webcam, Facebook, Instagram, YouTube & 4.0 & Technology & Google, Maps, button, lite, free, safe, dark \\
& 27.7 & Violence & violent, killer, arrest, shooting, suicide & 2.3 & Years & 1977, 1972, 1965, 1963, 1923, 1940, 1905, 1857 \\
& 18.2 & Drugs \& Rebellion & drugs, alcohol, weed, rebel, band, DUI & & \\
& 5.3 & Sex & sex, nude, porn, breasts, lust, virgin, panties & & \\
\hline
\multirow{3}{*}{GloVe} & 18.1 & Sex & sex, erotic, orgasm, porn, kinky, incest & 21.3 & Infrastructure & infotech, grid, construction, metro, railway \\
& 13.9 & Sex (Headlines) & Sexy, Naked, BDSM, Teenage, Babes, Lesbian & 44.8 & Politics & Dharmashala, Al Qaeda, anti-government \\
& 9.0 & Violence & violent, torture, suspects, felony, rape & .01 & Music & mixing, mastering \\
& 29.8 & Technology & cellphone, cyber, clicks, streaming, risky, manga & 1.0 & Entertainment & Pathao, Tootle, Cartoonz, corporation, heroes \\
& 29.2 & Celebrity Names & Rihanna, Spears, Olson, Lindsay, Megan, MTV & 32.7 & Sports Names & Baniya, Neupane, Dhoni, Ashutosh \\
\bottomrule
\end{tabular}
\caption{\footnotesize Clusters of the most and exclusively associated words with the Teenager group in English and Nepali embeddings.}
\label{tab:cluster_table}
\end{table*}

\section{Results}

Results show biases in SWEs and GLMs reflective of the traditional media sources on which they trained. Data from workshops shows AI is misaligned with adolescent life, and adolescents are themselves aware of media biases.

\subsection{Static Word Embeddings}

Table \ref{tab:cluster_table} illustrates teenage life in clusters of words most-associated and uniquely associated with adolescents. Some clusters are descriptive, with words that mean \textit{teenager}, words related to school, common names of teenagers, and words for adjacent concepts like other age groups (\textit{baby-boomers}). We derived four themes from SWEs.

\subsubsection{Instability and Stereotypes} Among the most associated words in English SWEs, we find clusters of stereotypical descriptions (\textit{acne-ridden, braces-wearing, spiky-haired}), media stereotypes (\textit{jocks, nerd, emo, punks}), and words connoting mental illness (\textit{self-esteem, psychotic, suicidal}). A teenage rebellion cluster further illustrates the extent to which adolescents are seen as not in control of their desires, with words such as \textit{sex-crazed} and \textit{drug-crazed}. A similar Drugs \& Rebellion cluster forms among the uniquely associated FastText words, highlighting teen use of drugs and alcohol. These associations find little analogue in Nepali SWEs, as we do not observe comparable associations with stereotypes and instability. \looseness=-1

\subsubsection{Violence and Vulnerability} Risk and violence emerge in the English SWEs. Words like \textit{victim} and \textit{at-risk} indicate teenage vulnerability to violence, while \textit{killer} and \textit{suspects} suggest teenagers as perpetrators. Violence takes forms from bullying, to lethal violence such as \textit{murder} and \textit{suicide}, to sexual violence including \textit{rape}, to criminal violence (\textit{arrest}, \textit{felony}), to sensationalized violence like \textit{torture}. Violence composed the single largest cluster of uniquely associated words (27.6\%) in the English Fasttext SWE. We identified a Violence cluster in the most associated Nepali GloVe words (\textit{fugitive, murder, police}), but it is notably smaller than English Violence clusters, and mostly free of sensationalized violence.

\subsubsection{Sex and Sexualization}

Sexual taboo and fetishization of adolescents emerge in the most and uniquely associated words in English SWEs. Words like \textit{lolita, underage, barely-legal}, and \textit{jail-bait} occur in the most-associated words, along with \textit{voyeur}. The word \textit{porn} occurs among uniquely associated words, along with a cluster of capitalized words including (\textit{BDSM, Lesbian, Naked}), suggesting an origin in the headlines of pornographic webpages. Pornographic and fetishizing clusters are distinct from clusters of sexual desire words, which occur in Nepali and English SWEs and include words like \textit{lust, sexual pleasure,} and \textit{lovers}.

\begin{figure*}[!htbp]
\centering
\begin{tikzpicture}[baseline]
    \begin{axis} [
        height=4cm,
        width=\textwidth,
        ybar = .05cm,
        bar width = 10.5pt,
        ymin = 0, 
        ymax = 35,
        ylabel=\% of Continuations,
        ylabel style={font=\footnotesize},
        ylabel shift=-2pt,
        xtick = {1,2,3,4,5,6,7,8},
        xtick style={draw=none},
        ytick pos = left,
        ymajorgrids=true,
        xticklabels = {Social Problems, Relationships, Teen Experiences, Law and Crime, Age, Media and Culture, Politics, Education},
        title=Deductive Codes Applied to GLM Prompt Continuations,
        xticklabel style={font=\scriptsize},
        enlarge x limits={abs=1cm},
        legend style={at={(0.81,0.57)},
        anchor=south west, 
        legend columns=1,
        font=\footnotesize},
        title style={font=\footnotesize},
    ]    
    \addplot [pattern=crosshatch, pattern color = red] coordinates {(1,30.22222222222222) (2,25.77777777777778) (3,20.88888888888889) (4,20.88888888888889) (5,9.333333333333334) (6,8) (7,6.222222222222222) (8,6.222222222222222)};    
    \addplot [pattern=vertical lines, pattern color = blue] coordinates {(1,29.333333333333333) (2,33.33333333333333) (3,19.555555555555557) (4,10.666666666666667) (5,0.0) (6,4) (7,0.8888888888888889) (8,13.777777777777778)};    
    \addplot [pattern=horizontal lines, pattern color = green] coordinates {(1,13.333333333333333) (2,17.333333333333333) (3,23.55555555555556) (4,2.2222222222222223) (5,2.666666666666667) (6,4) (7,17.777777777777778) (8,10.222222222222223)};    
    \legend {GPT-2-XL EN, LLaMA-2-7B EN, distilGPT-2 NE};    
    \end{axis}
    \end{tikzpicture}
    \caption{\footnotesize Social problems predominate in the English-language generative models continuing prompts related to teenagers. }
    \label{fig:lmcodes}
\end{figure*}
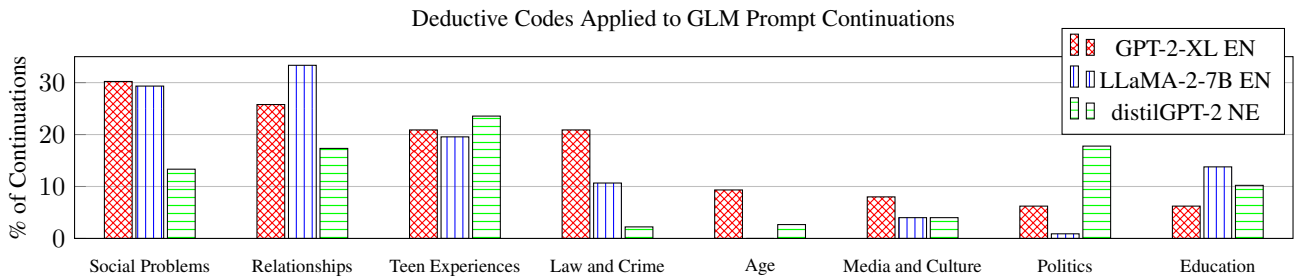

\subsubsection{Emerging Adulthood} The English FastText SWE includes a Coming-of-Age cluster (\textit{coming-of-age, right-of-passage}), while clusters related to the bodily transition of puberty occur in English SWEs (\textit{puberty, gawkiness}) and Nepali SWEs (\textit{puberty, menstruation}). The Nepali FastText cluster also includes words related to taking on adult roles in marriage and work (\textit{marriageable, employable}). Moreover, though we did not appreciate it until interacting with Nepalese adolescents, Infrastructure (\textit{infotech, construction} and Public Events (\textit{demonstration, program}) clusters also point to emerging adulthood, as adolescents can graduate from high school after the equivalent of the 10th grade, and can take a job in a trade, beginning adult life.

\subsection{Generative Language Models}

Figure \ref{fig:lmcodes} visualizes the deductive codes applied to the 225 continuations from LLaMA-2-7B, GPT2-XL, and distilGPT2-Nepali, based on the prompts in Table \ref{tab:lm_prompts}. We derived three themes for GLMs.

\subsubsection{Social Problems - Especially Violence - Are Common}

30\% of GPT-2-XL continuations and 29\% of LLaMA-2 continuations received the Social Problems code, making this the most common code for GPT-2-XL and the second most common for LLaMA-2. Of the Social Problems continuations, 47\% were subcoded for Violence in LLaMA-2, and 50\% in GPT-2-XL. For example, the following was generated by LLaMA-2 from ``At home, the teenager'': \textit{was bullied by his mother's boyfriend. At school, he was taunted by the kids. He was so depressed, he attempted suicide.} Other common subcodes included Drug Use (21\% LLaMA-2, 9\% GPT2-XL); Teen Trauma (17\% LLaMA-2, 21\% GPT2-XL); Mental Illness (9\% LLaMA-2, 12\% GPT2-XL); and Sexualization (9\% LLaMA-2, 13\% GPT2-XL), as in this continuation from GPT2-XL: ``Online, the teenager'': \textit{was charged with child porn and illegal computer access. After the investigation was closed into his alleged illegal access, a case had to be filed.} Though much less common, violence also occurs in the continuations of DistilGPT2-Nepali. Bullying is absent, but suicide and sexual violence occur in the roughly 2\% of continuations coded as Law and Crime. Though social problems are the default in English, we also observe teenage exemplars - exceptions who are noteworthy for deviating from the norm. For example, LLaMA-2 continues ``At school, the teenager'' with \textit{has a very good academic record, and is a member of the student council. In addition to her school duties, she has been a member of the Girl Scouts since she was in the first grade.}

\subsubsection{Sensationalism Emerges from ``High-Quality'' Training Data}

Many GLM continuations, including those resulting in social problems and violence, either 1) followed a distinct journalistic style or 2)  explicitly cited a news media source or described a quote being taken by a media source. The following representative example from LLaMA-2-7B was generated from ``At school, the teenager'': \textit{was bullied for his sexual orientation. The 15-year-old boy from the village of Nizhny Novgorod, who was bullied for his sexual orientation, committed suicide.} The continuation follows a journalistic style that concisely communicates the boy's age, hometown, and circumstances leading to the events under consideration. In other cases, the model appears to shift into a journalistic mode of writing; LLaMA-2 continues ``The teenager flirted because they wanted: \textit{to have sex with her. A 17-year-old girl from Warrington has been found guilty of having sex with a 14-year-old boy.} Other continuations identify quotes taken by media outlets, including CNNMoney, KRIV-TV (an NBC affiliate, according to GPT2-XL), and the Daily News. In one case, a LLaMA-2 output noted that photos were provided by Getty Images. Continuations by DistilGPT2-Nepali often included the apparent source of the model's continuation, such as Everest Online News, eHimala, Today's News Media Prof, and the Federation of Nepal Journalists. This suggests even models trained on reputable sources of text data are nonetheless vulnerable to sensationalism and societal bias, if reflected in the press.
 
\subsubsection{Societally Sanctioned Activities for Adolescents}

The codes appropriate to GLM continuations also surfaced societal attitudes toward specific adolescent activities. Prompts involving parties were the most likely to result in continuations involving social problems, followed by prompts involving teenagers online. Prompts involving teenagers in the workplace, on the other hand, were the least likely to produce continuations involving societal problems, even if many English-language continuations do trivialize adolescent experiences at work, as in the case of several LLaMA-2 continuations involving adolescents being fired because they refused to take drug tests. Prompts involving school were the most likely to be coded for adolescent relationships, while prompts involving the home were the most likely to involve adolescent experiences, as in the LLaMA-2-7B continuation of ``At home, the teenager'': \textit{is a person who is looking for their identity. They are trying to find out what they are about.} Finally, prompts involving adolescents online were the most likely to result in continuations related to media and culture.

\begin{table*}
\fontsize{7}{7.5}\selectfont
\centering
\begin{tabular}{ccc||ccc}
\toprule
 \multicolumn{3}{c||}{\textbf{\textit{Most}} Similar Words (U.S. Participants)} & \multicolumn{3}{c}{\textbf{\textit{Most}} Similar Words (Nepalese Participants)} \\
 \hline
 \% & \textbf{Cluster Name} & \textbf{Representative Words}  & \% & \textbf{Cluster Name} & \textbf{Representative Words} \\
\hline
10.7 & Fun & fun, party, fashion, curiosity &  23.5 & Energy & energetic, playful, excited, emotional \\
 12.0 & Stress & stress, moody, rebellious, reactive & 26.5 & Stress & stress, pressure, fear, gossip, angry \\
 12.0 & Immaturity & immature, irresponsible, insecure, anxiety & 10.3 & Immaturity & immaturity, shy, ignorant, fake \\
 20.0 & Discovery & discovery, growth, independence, identity & 7.4 & Innocence & childhood, innocent, obedient, sleepy \\
 20.0 & Social Life & social, friendly, family, bonds & 32.4 & Likability & friendly, cool, beautiful, youth \\
 12.0 & School & grades, homework, procrastination, curious & & & \\
 8.0 & Boredom & bored, lazy, dull, tired & & & \\
 5.3 & Difference & different, makeup, sleep, phone & & & \\
\hline
 \multicolumn{3}{c||}{\textbf{\textit{Exclusively}} Similar Words (U.S. Participants)} & \multicolumn{3}{c}{\textbf{\textit{Exclusively}} Similar Words (Nepalese Participants)} \\
 \hline
 \% & \textbf{Cluster Name} & \textbf{Representative Words}  & \% & \textbf{Cluster Name} & \textbf{Representative Words} \\
\hline
 18.9 & Uncertainty & questioning, overthinking, impulsive, ambitious  & 15.3 & Pressure & pressure, showoff, drama, ruthless \\
 26.4 & Change & changing, different, curious, frisky & 20.8 & Freedom & freedom, independent, dynamic, creative \\
15.1 & Impatience & impatient, restless, reckless, moody & 19.4 & Impatience & restless, irritation, unsatisfied, greedy \\
22.6 & Inexperience & confused, misunderstood, inexperienced, gullible & 8.3 & Inexperience & uninformed, shy, lazy, solitary \\
 17.0 & Eagerness & idealistic, impressionable, attentive, college & 9.7 & Adventure & adventurous, excited, expressive, emotional \\
 & & &  16.7 & Likability & chill, clever, fashionable, good \\
 & & &  20.8 & Discipline & disciplined, work, study, attitude \\
\bottomrule
\end{tabular}
\caption{\footnotesize Clusters of most and exclusively associated words describing teenagers, according to teen participants in the U.S. and Nepal.}
\label{tab:human_cluster_table}
\end{table*}

\subsection{Workshop Sessions}

Workshop data demonstrates that AI reflections of teenage life are disconnected from the experiences of adolescents. We derived three themes from participant responses.

\subsubsection{AI Does Not Reflect Adolescent Views of Adolescence} 

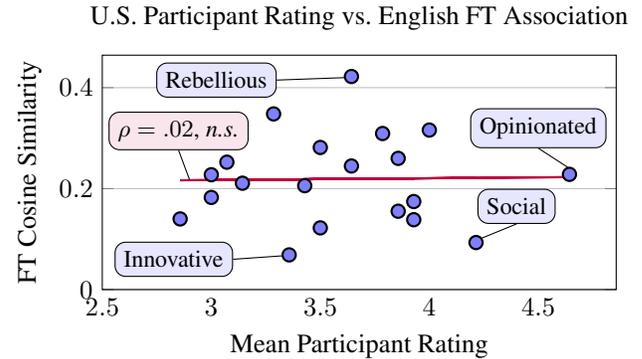
\begin{figure}
\begin{tikzpicture}
\centering
\begin{axis}[
legend pos=outer north east,
width=\linewidth,
height=4.7cm,
ymin=0,
xmin=2.5,
ymajorgrids=true,
ylabel={FT Cosine Similarity},
xlabel={Mean Participant Rating},
xtick style={draw=none},
title={U.S. Participant Rating vs. English FT Association},
ylabel near ticks,
ylabel shift=-.2,
legend style={
    at={(0.14,.54)}, 
    anchor=south,
    legend image post style={scale=2},
    }
]
\addplot [only marks, mark = *, mark size=2.5, mark options={color=blue!50, draw=black, opacity=0.5, thick}] table[x=humanMeans, y=modelSims] {\fttable};
\addplot [thick, purple] table[
    x=humanMeans,
    y={create col/linear regression={y=modelSims}}
]
{\fttable};

\draw[-] (3.62,.42) -- (3.3,.415) node[left, draw, fill=blue!10, rounded corners] {\small Rebellious};

\draw[-] (4.65,.238) -- (4.5,.28) node[above, draw, fill=blue!10, rounded corners] {\small Opinionated};

\draw[-] (4.23,.1) -- (4.4,.13) node[above, draw, fill=blue!10, rounded corners] {\small Social};

\draw[-] (3.33,.067) -- (3.1,.06) node[left, draw, fill=blue!10, rounded corners] {\small Innovative};

\draw[-] (2.9,.218) -- (2.85,.276) node[above, draw, fill=purple!10, rounded corners] {\small $\rho=.02$, \textit{n.s.}};

\end{axis}
\end{tikzpicture}
\caption{\footnotesize Word associations with ``teenager'' in FastText are decorrelated from U.S. teens' ratings of their similarity to ``teenager.''}
\label{fig:teen_crawl}
\end{figure}

As discussed in the Methods, participants rated 20 trait words (e.g., \textit{opinionated, thoughtful}) from 1 to 5 based on how well they described teenagers. We took the same words and computed the cosine similarity between the $\vec{teenager}$ vector and the trait word vector. We then took the correlation between mean participant ratings and cosine similarities, obtaining Pearson's $\rho$=.02, \textit{n.s.} for English FastText, and $\rho$=.06, \textit{n.s.} in English GloVe, indicating no correlation between SWEs and human ratings, as shown in Fig \ref{fig:teen_crawl}. Similar results were obtained for Nepali embeddings, with $\rho$=.06, \textit{n.s.} in Nepali FastText, and $\rho$=$-$.23, \textit{n.s.} in Nepali GloVe.

As shown in Table \ref{tab:human_cluster_table}, we also clustered the most-associated and uniquely-associated words provided by teenagers, using a vector for each word based on its valence, arousal, and dominance in the lexicon of \citet{mohammad2018obtaining}, and applying the k-means algorithm. U.S. clusters suggest a strikingly different view of adolescent life than that of English SWEs. Clusters related to School, Social Life, Discovery, and Fun  make up more than 60\% of the clustered most similar words. Where more negative traits like \textit{rebellious} and \textit{insecure} emerge, they are balanced by apparent explanations suggested by words like \textit{stress} and \textit{anxiety}. Clusters of exclusively associated words bear more resemblance to English SWEs, with Change and Uncertainty making up more than 45\% of the clustered words. However, the clusters also surface feelings of Inexperience (\textit{confused, misunderstood, gullible}) and Eagerness for the future (\textit{idealistic, attentive, college}). Notably absent is \textit{any} word connoting violence or lurid sexuality. Nepalese exclusively associated clusters similarly describe Impatience, Inexperience, and interest in Freedom and Adventure. Clusters related to Likability (\textit{cool, beautiful, chill, clever}) occur in both the most and exclusively associated words, while words related to Pressure and Discipline, with a particular focus on school (\textit{disciplined, study, pressure}), make up more than 35\% of the clustered exclusively associated words.

\subsubsection{Adolescent Life is Not Well-Characterized by Newsworthy Events}

Qualitative analysis showed that participant prompt continuations were misaligned with the continuations of GLMs. Prompted with ``At school, the teenager'', U.S. participants responded with \textit{writes in a notebook} (E8), \textit{doesn't pay attention to the teacher} (E1), \textit{studies in class} (E12), and \textit{eats lunch} (E5). Prompted with ``At home, the teenager'', four U.S. participants wrote about videogames, three about sleeping, and two about homework. Videogames and watching online videos on platforms like Tiktok also constituted the majority of responses to the prompt ``Online, the teenager.'' Six continuations of ``At the party, the teenager'' included talking to friends, while two discussed drinking alcohol. Aside from one mention each of cyberbullying and shoplifting, participant continuations are devoid of violence, rebellion, and sexualization. A far cry from the social problems in GLMs, the only description of a teenager facing discipline is specified by E10 for ``With their teachers, the teenager'': \textit{got in trouble for sleeping in class.}

Responses from Nepalese participants were similarly mundane. Continuing ``At school, the teenager,'' nine participants described studying, learning, or reading, two described respecting teachers, and two described getting scoldings or beatings from teachers. In response to ``At home, the teenager'', five participants described doing chores, three using a cellphone, two browsing social media, and three doing homework. In response to ``At the party, the teenager,'' five participants described dancing, three wearing new or beautiful clothing, and three eating or feasting. In response to ``Online, the teenager'', six participants described searching for information or studying, five chatting or gossiping, and two playing games. Far from the sensationlized outputs of SWEs and GLMs, adolescents describe everyday activities: going to school, playing videogames, and talking with friends.

\subsubsection{Societal Expectations Inform Adolescent Presentations of Adulthood} Comparing responses of U.S. and Nepalese participants revealed differing manifestations of emerging adulthood. Responding to ``At work, the teenager'', eight Nepalese participants wrote that the teenager is hardworking, while three others described focusing, or being fired due to lack of focus. In response to ``The teenager worked because they wanted'', seven participants described a \textit{shortage} or need of money, and two more described helping with family finances. By contrast, every U.S. participant wrote \textit{money}, describing potential uses of this money to buy clothes (E9), new games (E10), a car (E11), or just \textit{stuff} (E1, E12). E3 wrote \textit{the freedom that money allows while having minimal bills.} Responding to ``At work, the teenager'', three U.S. participants described completing assigned tasks, two talking to friends or coworkers, playing on their phone (E11), ignoring their manager (E13), or doing the \textit{bare minimum} (E1). Where U.S. participants described work as an avenue to independence and agency, Nepalese participants described it as a means of supporting their family. Both descriptions reflect emerging adulthood, contextualized by the expectations and opportunities of two societies.

\subsection{Instructions for Fair AI}

Participants wrote instructions for AI to represent teenagers fairly, and shared thoughts on the sources of data on which AI trained. We arrived at four themes based on this data.

\subsubsection{Adolescents are Aware of Media Stereotypes}

U.S. participants contended that media representations of teenagers are biased and reflect a stigma around adolescence. E7 wrote: \textit{Out of all age groups, teenagers are by far the most stigmatized} and \textit{many people hold stereotypical views of teenagers\dots consistently reinforced through media.} Similarly, E4 wrote \textit{teenagers are viewed in a very negative light because we have a tendency to deal with things in a very different way than adults or people from other age groups deal with their problems.} Nepalese participants also highlighted that societal views differ from those of teenagers. N16 wrote that it is \textit{important to describe the teenager as they are\dots teenagers' views are different from society's point of view.} N13 wrote \textit{teenager[s] aren't like the society think[s,] because they create their own way.} Participants also noted that \textit{how} AI learned about adolescents would affect their view of using it. E8 wrote: \textit{for teenagers to feel seen or heard I think it would be good to have them be the ones that tell [AI] about themselves and not have [it] assuming.} E6 wrote that, were AI to train on \textit{data on teenagers from the media, [it] would most likely learn what a stereotypical teenager is like and not how they actually are. The media usually puts teenagers in a bad light but\dots they can be smart, well mannered, and successful.} E10 wrote that AI trained on media would be disconnected from teenage life, noting \textit{Teens make fun of how movies and TV shows portray them, finding it to be really far off from what they are in real life.} Finally, N13 wrote \textit{AI should represent [teenagers] as they are rather than what other[s] think of them.}

\subsubsection{No Media Source is Unbiased, But Some are More Biased Than Others}

Reflecting on using traditional and online media sources for AI training data, E11 wrote: \textit{movies, newspapers, and other media often portray teens in a stereotypical fashion that only captures part of what a teen really is. The information\dots would be surface level at best.}  E13 wrote that if AI systems \textit{read the newspaper, much of the information they would gain could be false as it is the way others view teenagers rather than the way they actually are. Whereas teenagers would be able to provide the real way they see themselves.} N4 stressed the disconnect between media and reality, writing \textit{what we learn from media and newspapers is different [from] when we learn from human beings.}

Participants acknowledged that perfectly unbiased media might be unachievable. E1 wrote: \textit{I think it is almost impossible to represent teenagers, or anything really, in media without some kind of bias.} E9 further noted: \textit{the way social media represents teenagers can be very far-fetched, and possibly even offensive to what teenagers are really like. I believe it’s important for\dots AI to accurately represent teenager[s] in comparison to possible lies and fake information being spread about them. But\dots all teenagers are different so I don’t believe there’s a specific way to represent them all accurately.} E3 highlighted that the attention-driven business model of media companies underlies the problem, writing \textit{I don't think the media is a good representation of any group of people because of the business model they work under.}

Most participants agreed that AI should interact with teenagers to learn about them. N17 wrote: \textit{Teens know more about themself than [any] other. So if teenagers teach [AI] about them it will be more effective compare[d] to learning about them from other media.} N1 wrote: \textit{media only explains about surface feeling[s,] but a teenager could explain about it in detail.} Finally, E10 suggested that AI might \textit{search through past chats with other teens in order to figure out what shared interests most teenagers have,} a strategy similar to that employed by many chat-based language models, which train on datasets of conversations \cite{zheng2023lmsyschat1m}. While such a dataset might raise an array of ethical concerns, E10 identifies a gap in training data for conversational models specific to underrepresented user groups.

\subsubsection{Diversity and Positivity: Perspectives on Fair Representation}

Two perspectives on how adolescents could be fairly represented by AI emerged in the data. U.S. participants (nine of thirteen) stressed portraying the \textit{diversity} of teenagers. E7 wrote: \textit{Instruction 1: Clarify that not all teenagers are the same. As it is with every age group, traits can vary drastically between individuals.} E3 wanted to ensure that AI would \textit{include examples of teenagers from all backgrounds.} E9 noted: \textit{teenagers are all very different\dots there’s no specific category to place teenagers under.} The preference for diverse representation was sometimes juxtaposed with an assumption that AI would focus on adolescents' negative traits. E1 wrote: \textit{Instruction 1: When asked about teenagers, don’t just say the bad things; teenagers are different from each other, so you should represent all of them.} E13 wrote: \textit{Give both good and bad examples. For example, mention that they are rebellious but also innovative.} Where U.S. participants stressed diversity, Nepalese participants centered \textit{positivity}, with ten participants listing positive traits in instructions to AI. N9 wrote that AI should reflect that \textit{teenagers are the most creative and confiden[t] and thoughtful.} N13 similarly wrote that teenagers \textit{are free minded, introvert[ed], and curious.} While the preference for diversity may reflect a U.S. cultural value, the motivation is similar between U.S. and Nepalese participants: to present adolescents generously, including positive traits rather than replicating negative media biases.

\subsubsection{The Potential for AI to Correct Stereotypes}

Both U.S. and Nepalese participants expressed optimism that AI could help in correcting stereotypes. E10 positioned AI as a mediator, writing that \textit{society has a negative stereotype of teenagers, that they are moody for no reason and that they are disrespectful. But teens have various reasons for acting the way they do, and [AI] could help people understand that.} E13 suggested proactively addressing biases, writing \textit{there is no way to break the social stereotype that teenagers act a certain way if the only information being put out about teens supports the stereotype, rather than showing the stereotype is false.} N4 wrote that \textit{AI could express the teenagers in [a] way [that] every one will accept it.} Highlighting that AI could serve as a vector for better interpersonal communication, N7 said that \textit{society should also know about how the teenagers feel and the way they think.} In contrast with existing information architectures like social media, N1 wrote that \textit{AI could be the place where teenagers feels safe.}
\section{Discussion}

Our work shows that even training on high-quality data sources, such as newspaper articles, can reproduce harmful societal attitudes that depict adolescents as violent, criminal, and rebellious. That some of these biases - particularly related to violence - do not exist in monolingual Nepali-language models might prompt us to re-examine assumptions that these biases are unavoidable. That more user-facing generative models reflect association of adolescence with social problems shows the potential for AI to amplify biases, as GLMs begin to serve as mediators of culture \cite{brinkmann2023machine,dangol2024mediating} and sources of information \cite{memon2024search}.

\subsection{Epistemic Infrastructure for Adolescents}

Adolescents' access to information and shared spaces is often mediated by societal attitudes. For example, \citet{bernier2011representations} find that only 2.2\% of facility square footage is devoted to teenage users in libraries, where youth represent nearly 25\% of all users, observing that this disparity is motivated by unsavory stereotypes about adolescents and serves to marginalize them in an essential space for information seeking. As AI begins to serve society's information seeking needs, our work poses the question of whether AI can serve as a \textit{place where teenagers feel safe}, as N1 put it, or if it will reflect the attitudes and serve primarily the needs of adult users. Feeling safe using AI might also support teen development by providing a space to ``enact maturity,'' inviting adolescents into conversations about consequential subjects, like politics \cite{ballard2022opportunities}.

\subsection{Addressing Societal Bias with AI}

Participants saw AI as a means of addressing societal stigma in traditional and social media. To do so, they believed AI would need to understand adolescents by interacting directly with them, rather than reading about them in secondary sources. Some participants even envisioned AI mediating between adolescents and adults, providing perspective when teens aren't able to express themselves. Such optimism about the role of AI suggests the need to develop frameworks for ethical engagement between adolescents and language technologies. While AI may hold potential for changing societal attitudes toward teenagers, it can also be used to collect data or financial resources from users \cite{wolfe2024expertise}. Finding ways to maximize user agency while personalizing models could be explored in future work. \looseness=-1

\subsection{Human Perspectives in Studies of AI and Society}

Our study paired an analysis of a societal attitude in AI with a human subjects study of the group impacted, revealing the disconnect between adolescent experiences of the world and AI presentations. Participants provided context that allowed us to understand how societal expectations of teenagers shape their self-presentation, and their presentation in media sources. Our work indicates that more complete descriptions of AI and societal biases can be obtained through mixed methods work, involving not only AI-based measurements but also participation of human subjects.

\subsection{Limitations and Future Work}

We used solely monolingual, open models to maximize reproducibility and prevent cross-lingual transfer of semantic associations. Nonetheless, we acknowledge that most users prefer proprietary, chat-based, multilingual models like ChatGPT. Future work might examine such models not as representative reflections of culture but as sociotechnical tools, considering how they affect the lives of adolescents. Moreover, while the Nepali-language models used are the best we know of, we nonetheless observed some disfluencies in their output, a continuing limitation of low-resource languages.  Finally, adolescents are a group so large and diverse that we cannot hope to fully capture them in a single study. We hope that future research will consider additional populations of adolescents in the U.S. and around the world.
\section{Conclusion}

We showed that the often lurid and sensationalized depictions of adolescents present in AI are decoupled from the everyday experiences of U.S. and Nepalese adolescents, whom our workshops revealed are well-aware of media stereotypes. Even as teenagers grapple with perceived social stigma, they view AI as having potential to effect positive change, establishing a safer and more positive environment for adolescents. We hope this research will inspire work that will realize that goal, and will provide a starting point for future studies of bias that not only draw on AI reflections of society but also surface the perspectives of those affected.

\section{Ethical Considerations}

While we believe our participant samples to be representative, we do not intend by studying under-represented groups in the present work to flatten or essentialize the experiences of these individuals. We note in the paper that societal understandings of adolescence have changed over time, and our participants noted that individual experiences can vary widely regardless of membership in a given demographic group. We also note that, despite participants' enthusiasm for AI to mitigate biases, leveraging new technologies to address societal problems comes with significant uncertainty, and future work is needed to study the efficacy, impact, and potential adverse impacts of such interventions. 

\section{Researcher Positionality}

Our work considers the perspectives of adolescents residing in the United States and Nepal. We have sought to accurately and fairly represent the opinions of these individuals, though we acknowledge that our positionality is necessarily limited in that all four authors of the present work are over the age of 18 and under the age of 65, meaning that we would belong to the Adults age group according to the NIH \cite{NIHages}. Moreover, all four authors currently reside in the United States. However, one of the first authors was born in Kathmandu, Nepal and resided there until the age of 18. In the time since, she has maintained relationships with schools in Nepal, and introduced culturally responsive computing education curricula for Nepalese learners. With respect to research background, two of the authors of the present work have extensive backgrounds in machine learning, specifically in studies of bias and fairness in artificial intelligence. The other two authors have extensive backgrounds in HCI and computing education, including AI for children and adolescents.

\section{Adverse Impacts}

We caution against readings of our work that would produce moral alarmism of the kind that resulted in sensationalized portrayals of adolescents in AI. We suggest that an appropriate response to this research is to consider not whether and in what situations adolescents should have access to AI, as in many discussions of adolescents and technology \cite{stern2017constructing}, but to consider the implications of how misaligned the societal discourse surrounding adolescents is from their lived experiences.

\section{Acknowledgements}

This project was supported in part by the Connecting the EdTech Research EcoSystem (CERES) research network.

\bibliography{references}

\appendix

\section{Appendix}

\subsection{Nepali GloVe Training Hyperparameters}

Note that we saved checkpoints and gradients during training to prevent the loss of data in the event of failure. Setting min vocab count to 5 prevents words that do not occur at least five times in the training corpus from appearing in the embedding, reducing semantic noise in the embedding.

\begin{table}[h]
    \centering
    \begin{tabular}{c|c}
    \hline
        Parameter & Value  \\
    \hline
    Min Vocab Count & 5 \\
    Vector Size & 300 \\
    Max Iter & 20 \\
    Window Size & 10 \\
    X-Max & 100 \\
    \hline
    \end{tabular}
    \caption{Hyperparameters for training Nepali GloVe model.}
    \label{tab:glove_training_params}
\end{table}

\subsection{Nepali Generative Language Model}

We specifically used the DistilGPT2-Nepali model available via the HuggingFace Hub at \url{https://huggingface.co/Sakonii/distilgpt2-nepali}. This model is trained on the nepalitext dataset, also available via the Hub at \url{https://huggingface.co/datasets/Sakonii/nepalitext-language-model-dataset}.

\subsection{Notes on Monolingual Training Corpora}

Training corpora for English-language SWEs and GLMs may contain some non-English text, but this is more by accident than design; 89.7\% of LLaMA-2 training data is in English, compared to less than .2\% for any other language \cite{touvron2023llama2}. More common due to the prevalence of English on the internet is the presence of English-language text in training corpora for low-resource SWEs. We observed that a portion of the exclusively associated Nepali-language words, mostly those related to technology and online applications, were English-language words. This likely reflects the use of code-switching by Nepali speakers, especially when discussing subjects such as western technologies.

\subsection{Notes on Decoding Parameters}

Note that we use the default for the HuggingFace implementations of GPT2-XL and DistilGPT2-Nepali, which is to set the max-length parameter to 50. For the quantized LLaMA-2-7B model, we allow the model to generate between 25 and 50 new tokens using the max-new-tokens and min-new-tokens parameters, as we found that the model sometimes failed to generate text using only the max-length parameter. Note that the HuggingFace library by default sets the \textit{top-k} parameter to 50, meaning that the models sample from the 50 highest-probability tokens at each decoding step. Our strategy, which examines the \textit{output} of a GLM, rather than its intrinsic properties, is similar to that of \citet{sheng2019woman}, who study the regard in which GLMs hold various demographic groups. For more information about decoding methods, multinomial sampling, and governing randomness using the temperature parameter, please see \url{https://huggingface.co/docs/transformers/generation_strategies#multinomial-sampling}.

\subsection{Notes on Human Subjects Word Clusters}

For human subjects word clustering, we could only cluster those words with valence, arousal, and dominance scores in the NRC-VAD lexicon \cite{mohammad2018obtaining}, meaning that we had to exclude a little over 20\% of English-language participant words. We also corrected misspellings and standardized forms using unusual punctuation such that a larger percentage of the submitted words could be clustered. The clusters in the paper are representative of the full set of words collected from participants, which we make publicly available in the supplementary materials.

\end{document}